\documentclass[fleqn,12pt]{wlscirep}
\usepackage{aas_macros}
\usepackage{setspace}
\usepackage{subcaption}
\usepackage{caption}

\usepackage{lineno}

\newcommand\xmm{\textit{XMM-Newton}}
\newcommand\chandra{\textit{Chandra}}

\newcommand\xrism{\textit{XRISM}}

\newcommand\delcstat{$\Delta$C-stat}
\newcommand\pcm{cm$^{-2}$}

\newcommand\logxi{$\log (\xi$/erg~cm~s$^{-1})$}

\title{Vertical wind structure in an X-ray binary revealed by a precessing accretion disk}

\author[1,*]{P. Kosec}
\author[1]{E. Kara}

\author[2]{A. C. Fabian}
\author[3]{F. Fürst}
\author[4]{C. Pinto}
\author[5]{I. Psaradaki}
\author[2]{C. S. Reynolds}
\author[1]{D. Rogantini}
\author[6]{D. J. Walton}

\author[7,8]{R. Ballhausen}
\author[1]{C. Canizares}
\author[9]{S. Dyda}
\author[10]{R. Staubert}
\author[11]{J. Wilms}

\affil[1]{MIT Kavli Institute for Astrophysics and Space Research, Cambridge, MA 02139, USA}
\affil[2]{Institute of Astronomy, Madingley Road, CB3 0HA Cambridge, UK}
\affil[3]{Quasar Science Resources SL for ESA, European Space Astronomy Centre (ESAC), Science Operations Departement, 28692 Villanueva de la Can\~nada, Madrid, Spain}
\affil[4]{INAF - IASF Palermo, Via U. La Malfa 153, I-90146 Palermo, Italy}
\affil[5]{University of Michigan, Dept. of Astronomy, 1085 S University Ave, Ann Arbor, MI 48109, USA}
\affil[6]{Centre for Astrophysics Research, University of Hertfordshire, College Lane, Hatfield AL10 9AB, UK}
\affil[7]{Department of Astronomy, University of Maryland, College Park, MD 20742, USA}
\affil[8]{NASA-GSFC/CRESST, Astrophysics Science Division, Greenbelt, MD 20771, USA}
\affil[9]{Department of Astronomy, University of Virginia, 530 McCormick Rd, Charlottesville, VA 22904, USA}
\affil[10]{Institut für Astronomie und Astrophysik, Universität Tübingen, Sand 1, 72076 Tübingen, Germany}
\affil[11]{Dr. Karl Remeis-Observatory and Erlangen Centre for Astroparticle Physics, Sternwartstr. 7, D-96049 Bamberg, Germany}

\affil[*]{pkosec@mit.edu}

\begin{abstract}

\textbf{The accretion of matter onto black holes and neutron stars often leads to the launching of outflows that can greatly affect the environments surrounding the compact object. An important means of studying these winds is through X-ray absorption line spectroscopy, which allows us to probe their properties along a single sightline, but usually provides little information about the global 3D wind structure that is vital for understanding the launching mechanism and total wind energy budget. Here, we study Hercules X-1, a nearly edge-on X-ray binary with a warped accretion disk precessing with a period of about 35 days. This disk precession results in changing sightlines towards the neutron star, through the ionized outflow. We perform time-resolved X-ray spectroscopy over the precession phase and detect a strong decrease in the wind column density by three orders of magnitude as our sightline progressively samples the wind at greater heights above the accretion disk. The wind becomes clumpier as it rises upwards and expands away from the neutron star. Modelling the warped disk shape, we create a 2D map of wind properties. This measurement of the vertical structure of an accretion disk wind allows direct comparisons to 3D global simulations to reveal the outflow launching mechanism.}

\end{abstract}

\begin{document}

\maketitle

Despite their existence being known for decades \cite{Tombesi+10, Ueda+04, Neilsen+09, Pinto+16}, the driving mechanism of accretion disk winds is still poorly understood. In supermassive black holes, these outflows can even be powerful enough to dictate the evolution of the entire host galaxy, and yet, to date, we do not understand whether they are launched by radiation pressure \cite{Shakura+73, Proga+00}, magnetic forces \cite{Blandford+82}, thermal irradiation \cite{Begelman+83}, or a combination thereof.

After the detection of an ionized outflow in its X-ray spectrum \cite{Kosec+20}, Hercules X-1 \cite{Giacconi+72, Tananbaum+72} (hereafter Her X-1) became the ideal object to study the physics of accretion disk winds. The regular precession of its warped disk \cite{Roberts+74, Peterson+75, Gerend+76, Katz+76} introduces a varying line of sight through the wind, allowing us to study its vertical properties ({\color{blue}Figure 1}). To follow-up this unique opportunity, we executed a large \xmm\ (380 ks exposure) and \chandra\ (50 ks exposure) campaign, performed in August 2020, and observed a significant portion of a single disk precession cycle of Her X-1. The campaign was composed of three long, back-to-back \xmm\ observations starting at the beginning of a new precession cycle - the \textit{Turn-on} moment, when our line of sight towards the neutron star is being uncovered by the tilted outer accretion disk, and Her X-1 enters into a \textit{High State} (see {\color{blue}Figure 1} for a schematic of the Her X-1 system).

We perform a time-resolved spectral analysis to study the variation of the disk wind properties over the precession phase by splitting the 3 long \xmm\ observations into 14 segments. Her X-1 is an X-ray binary system with very complex short-term as well as long-term behavior. To assess any long-term variations in the wind properties beyond a single precession cycle, we also utilize the rich \xmm\ archive (11 observations), and show the 2020 observations separately in case any variations occurred in the accretion flow. Finally, we also use 3 \chandra\ observations, 2 of which were simultaneous with the 2020 \xmm\ campaign.

\subsection*{The evolution of the disk wind properties with the precession phase}

The optical depth of the absorption features strongly decreases with the increasing precession phase. This variation is shown in \ref{xcamp_opt_depths} for the three consecutive \xmm\ observations from 2020, taken during a single precession cycle. While the absorption lines are very strong in the first observation (immediately after \textit{Turn-on}), they weaken in the second observation and are nearly absent during the third observation.

By applying photoionized absorption spectral models, we find that the line depth variation is due to a dramatic drop in the line-of-sight column density of the absorber, which decreases from $\sim2 \times 10^{23}$ \pcm\ at maximum to $\sim 2 \times 10^{20}$ \pcm\ towards the end of the \textit{Main High State} (phase$~\sim0.2$). The ionization parameter \logxi\ is also found to decrease with progressing precession phase, with the exception of the \textit{Turn-on} moment when it is relatively low at around \logxi\ of 3.0, and varies from about $4.0$ at maximum down to $2.5-3.0$ at the end of our coverage. The evolution of the column density and the ionization parameter with the precession phase is shown in Fig. \ref{nh_xi_suporb}, and listed in \ref{resulttable}. The observed variations are clearly driven by the precession phase and we do not find any correlations with Her X-1 luminosity or binary orbital phase (\ref{Evolution_nh} and \ref{Evolution_xi}). Importantly, we find that the August 2020 measurements are in general agreement with the archival dataset measurements, indicating limited long-term evolution of the disk wind structure. The outflow velocity at different sight lines (precession phases) is found to be $250-800$ km/s in most cases (consistent with ionized winds detected in other X-ray binaries \cite{Ponti+12}), and we find no systematic variation of the velocity with precession phase (\ref{zvcorr_vturb_suporb}).

The observed behaviour confirms our hypothesis that the apparent wind variation is connected to the warped disk precession, and it is introduced as the whole wind structure precesses along with the disk. Hence our sightline through the wind towards the X-ray source is changing with precession phase and we are sampling different parts of the outflow structure, at different heights above the disk. These results give us a unique insight into the global properties of an accretion disk wind, inaccessible in most accreting systems (without a known disk precession cycle) and directly show that the wind column density strongly decreases with increasing height above the disk. Such finding agrees with the results of statistical studies of black hole X-ray binaries \cite{Ponti+12}, where most wind detections were achieved in high inclination systems, suggesting an equatorial wind structure with a small opening angle (5-10$^{\circ}$).

The observations taken as part of this recent campaign and the archival \xmm\ and \chandra\ data all agree  on the rough shape of the observed trends, although we do observe a scatter in the best-fitting wind parameters. The scatter could be due to physical changes of parameters in a clumpy wind, but could also be attributed to our limited spectral resolution with \xmm\ in the crucial, but complex Fe K energy band between 6 and 7 keV (further discussed in Methods).

\subsection*{Absorber distance from the X-ray source}

We use the wind column density and ionization parameter to estimate the distance of the absorber from the X-ray source. The definition of column density is: $N_{\rm{H}}=n \Delta R=n R (\Delta R/R)$ and the definition of the ionization parameter: $\xi=L_{\rm{ion}}/(n R^{2})$, where $n$ is plasma density, $R$ the distance of the absorber, $\Delta R$ is the absolute and $\Delta R/R$ the relative thickness of the absorbing layer, and $L_{\rm{ion}}$ is the ionizing luminosity during each observation (between 13.6 eV and 13.6 keV). By combining the two expressions, we obtain: 

\begin{equation}
    R = \frac{L_{\rm{ion}}}{N_{\rm{H}}\xi} \frac{\Delta R}{R}
\end{equation}

The value of $\Delta R/R$ is unknown. To constrain the position of the outflow, we estimate both its maximum and minimum distance from the ionizing source. By setting $\Delta R/R=1$ since the relative thickness cannot be larger than unity, we calculate the maximum distance of the absorber from the X-ray source, shown in Fig. \ref{abs_dist} (top panel). At the beginning of the precession cycle, our observations sample the wind structure around the wind base, close to the outer accretion disk (which is moving away from our line of sight). Near the wind base the relative absorber thickness $\Delta R/R$ should be large, as the outflow is likely launched from the disk over a range of radii, and hence its position is close to the estimated maximum distance from the X-ray source. With the exception of the first two data points around phase $\sim0$, the wind structure at low heights is close to the thermal wind launch radius \cite{Begelman+83}, estimated to be about $8 \times 10^{9}$ cm in Her X-1 \cite{Kosec+20}.

The escape velocity at these radii is roughly 2000 km/s, higher than our line of sight wind velocity measurements ($250-800$ km/s). Thus the flow at these locations could be gravitationally bound, as suggested for Her X-1 and other X-ray binaries exhibiting blueshifted absorbers \cite{Nixon+20}. However, we note that the measured Doppler velocities are strict lower limits on the global (3D) outflow velocity. The wind could still be accelerating as it crosses our line of sight, and have significant toroidal (due to Keplerian motion) or vertical velocity components which cannot be measured from our nearly edge-on sightline.

The maximum absorber distance from the X-ray source then significantly increases at later precession phases, showing that the wind is likely expanding away from the accretion disk. If $\Delta R/R$ remains close to unity at later precession phases, the precessing wind structure spans a very large range of scales, extending to distances from the X-ray source of $10^{14}$ cm, significantly beyond the binary system separation (about $6\times10^{11}$ cm) \cite{Leahy+14}. The flight time of an outflow with a velocity of 500 km/s to $10^{14}$ cm is about 20 days, therefore it is improbable that the whole wind structure would be able to precess with the warped disk on precession timescales (the observed wind variation timescale is $\sim7$ days). Additionally, the assumption of $\Delta R/R \sim1$ at later phases results in unrealistically high wind mass outflow rates (calculated in Methods), much larger than the mass accretion rate through the outer accretion disk even after accounting for a limited wind launch solid angle.

Therefore $\Delta R/R$ must be significantly smaller than unity at later precession phases, and likely varies along the wind structure sampled with our changing line of sight. In other words, the wind fragments into smaller clumps. To recover meaningful wind physical properties, instead of $\Delta R/R \sim1$ , we can assume that the plasma density is constant. This may not be a completely correct assumption as the density is expected to drop with distance from the neutron star, however it will allow us to put at least a lower limit on the size of the wind structure, which may result in a more realistic size estimate than the first assumption of constant relative absorber thickness.

We take the wind measurement that results in the lowest absorber distance from the X-ray source based on the previous assumption ($\Delta R/R=1$), which likely exhibits the highest relative thickness of the absorbing layer (the relative thickness is likely to decrease with the distance from the X-ray source as the outflow expands and fragments). Using the definition of column density, we calculate the appropriate plasma density  $n$ for this observation, which is $3.6^{+4.1}_{-2.9} \times 10^{13}$ cm$^{-3}$ (archival \xmm\ observation 0673510501), similar to outflow densities measured in other X-ray binaries \cite{Psaradaki+18}. Finally, we calculate the absorber distances from the ionizing source for all other data points using this same density and the definition of the ionization parameter. The results are shown in Fig. \ref{abs_dist} (bottom panel).

Under the constant density, fragmenting outflow assumption, the absorber distance from the X-ray source changes much less at later phases and only reaches $5\times10^{10}$ cm. The wind location never reaches the outer disk boundary at $2\times 10^{11}$ cm, or at least our line of sight does not appear to cross such a part of the wind structure. The constant density assumption necessarily requires low relative thicknesses for the data points at later phases ($\Delta R/R\sim10^{-3}-10^{-4}$). In reality, both $n$ and $\Delta R/R$ may be variables (both decreasing with increasing distance from the X-ray source), creating a wind structure that extends between the two limiting cases illustrated in Fig. \ref{abs_dist}.

Finally, we notice that both data points at phase $\sim0$ are located farther away from the X-ray source in both plots in Fig. \ref{abs_dist} than the immediately following points at later phases. These parameter measurements are consistent in both archival and 2020 observations and indicate that  our line of sight may be sampling flows farther from the X-ray source at the very beginning of the disk precession cycle (just as the outer disk is uncovering our line of sight towards the neutron star). Alternatively, the material at these very low heights above the disk could simply be denser and more compact than at later times, but still located at the same physical position as the points at phase $\sim0.02$. This would be consistent with the material sampled at these phases still being located within the wind acceleration zone, where it is more compressed than at later times.

\subsection*{Vertical structure of the accretion disk wind}

By leveraging the estimated absorber distance from the X-ray source (under the constant density assumption) and modelling the shape of the warped Her X-1 accretion disk, we can measure the vertical position of the outflow. The disk is smoothly precessing with the 35-day period and thus for a given precession phase and radius in the disk, we can calculate the distance of the absorber from the disk itself (disk modelling is described in Methods). Repeating the calculation for all observations allows us to uniquely probe the vertical wind structure, and `X-ray' the accretion disk wind properties in 2 dimensions simultaneously.

The results are shown in Fig. \ref{constdist_location}, where the a 2D map shows the absorber position for all the analyzed observations, and the outflow properties (column density, ionization parameter) are shown using color scale. Most of the outflow locations are on a single apparent wind streamline, rising from the disk up to about $8\times10^{9}$ cm, moving away from the neutron star at an angle of $10-12^{\circ}$ with respect to the disk. We observe that both the wind column density and the ionization parameter strongly decrease along the streamline, as the wind expands away from the neutron star and fragments into smaller clumps. A similar variation is not observed for the outflow velocity, which appears to be uncorrelated with other relevant parameters.

These high-resolution X-ray spectra revealing the disk wind of Her X-1 open a new window into our understanding of the 3D structure of accretion disk winds in X-ray binaries. Making assumptions on the outflow density discussed above, we estimate the height and distance of the outflow from the X-ray source, but with next generation X-ray spectrometers, like \xrism, we may be able to observe density-sensitive absorption lines \cite{Miller+06} to test these assumptions. This 2D map of an accretion disk wind can be directly compared to theoretical simulations of X-ray binary accretion disk winds. In particular, the 3D properties of thermally and magnetically driven outflows could be compared with these observations, allowing us to understand which launching mechanism can explain better the outflow structure and physics seen in Her X-1, and in X-ray binaries in general. This unique approach to study the disk wind properties using a variable sightline may also be applied in a handful of other X-ray binaries thought to exhibit warped accretion disks, such as LMC X-4 \cite{Lang+81} and SMC X-1 \cite{Wojdovski+98}.

\section*{Acknowledgements}
PK thanks Michael Parker for helpful discussions about photoionization absorption spectral models in \textsc{xspec} and \textsc{spex} fitting packages. Support for this work was provided by the National Aeronautics and Space Administration through the Smithsonian Astrophysical Observatory (SAO) contract SV3-73016 to MIT for Support of the Chandra X-Ray Center and Science Instruments. PK and EK acknowledge support from NASA grants 80NSSC21K0872 and DD0-21125X. RB acknowledges support by NASA under Award Number 80GSFC21M0002. This work is based on observations obtained with \xmm, an ESA science mission funded by ESA Member States and USA (NASA). This research has also made use of data obtained from NASA’s \chandra\ mission.

\section*{Author contributions statement}

PK led the \xmm\ and \chandra\ observation proposal, spectral modelling and interpretation of results. EK contributed to the observation proposal, spectral analysis and interpretation of results. ACF, FF, CP, IP, CSR, DJW, RB, SD, JW contributed to the observation proposal and interpretation of results. RS contributed to scheduling the \xmm\ and \chandra\ observations and interpretation of results. DR and CC contributed to interpretation of results.

\section*{Data availability}
All \xmm\ data are available publicly at nxsa.esac.esa.int/nxsa-web. All \chandra\ data are available publicly at tgcat.mit.edu.

\section*{Code availability}

\xmm\ data were reduced using the \textsc{xmm sas} software and \chandra\ data were reduced using the \textsc{ciao} software. All spectra were analysed using the spectral fitting package \textsc{spex}. All figures except Fig. 1 were made in \textsc{veusz}, a Python-based scientific plotting package, developed by Jeremy Sanders and available at veusz.github.io.

\section*{Author Information} 
Reprints and permissions information is available at www.nature.com/reprints. The authors declare no competing financial interests. Correspondence and requests for materials should be addressed to PK (pkosec@mit.edu).

\bibliography{References}

\section*{Methods}

\subsection*{Data Preparation and Reduction}
\label{data}

Her X-1 has been observed many times with \xmm\ \cite{Jansen+01}. We analyze all 11 archival \textit{High State} observations (\ref{resulttable}), totalling 190 ks of raw exposure. However, the backbone of this paper is the Large campaign on Her X-1 carried out in August 2020, consisting of three back-to-back, full-orbit \xmm\ observations (380 ks raw exposure, listed in \ref{resulttable}). We also analyze 3 \chandra\ \cite{Weisskopf+02} observations (70 ks), out of which 2 were carried out simultaneously with the August 2020 \xmm\ coverage. These observations are described in more detail in sections 2 and 3 of a companion paper\cite{Kosec+22} (hereafter referred as Paper I). The three full-orbit \xmm\ observations were split into 14 segments for time-resolved spectral analysis. The segment exposure times were chosen to increase with the precession phase to offset the observed decrease in the disk wind absorption line strengths: the first observation was split into 7 segments ($\sim10$ ks duration each), the second into 5 segments ($\sim20$ ks duration), and the third into 2 segments (40 ks duration). In total we thus analyzed the spectra of 25 \xmm\ observations or segments and of 3 \chandra\ observations. 

We calculate Her X-1 orbital and precession phases for each exposure. The details of these calculations are provided in section 3 of Paper I. We determine the correction factor to any ionized outflow velocity measurements due to the neutron star orbital motion using Eq. 1 from our original Her X-1 study \cite{Kosec+20}. The formula uses a systemic velocity of the Her X-1 system of $-65$ km/s \cite{Reynolds+97}, and a projected orbital velocity of 169 km/s \cite{Deeter+81}. The effect of orbital eccentricity on the velocity correction factor is negligible.

For \xmm\ observations, we utilized data from the Reflection Grating Spectrometers \cite{denHerder+01} (RGS) in the 7 \AA\ (1.8 keV) to 35 \AA\ (0.35 keV) wavelength range, and data from the European Photon Imaging Camera (EPIC) pn instrument \cite{Struder+01} in the 7 \AA\ (1.8 keV) to 1.2 \AA\ (10 keV) range. For \chandra\ observations, we utilized the High Energy Transmission Gratings \cite{Canizares+05} in the range between 0.5 and 9 keV. Further information about our data reduction procedures is listed in section 3 of Paper I. 

EPIC pn was operated in Timing mode, but due to the extreme count rate of Her X-1, it was still piled-up in most observations. To counter pile-up, we progressively removed the central EPIC pn pixel rows until we reached residual pile-up of at most 5\% in the worst affected spectral bins (usually at the highest energies between 8 and 10 keV) in comparison with an even stricter extraction. We checked for the effect of this residual pile-up on our results using observations 0865440101 and 0865440401 by fitting pn spectra extracted with the stricter pile-up correction. The spectral fitting results were fully consistent with the results using the originally adopted pile-up correction (within errorbars), but naturally showed larger uncertainties due to the reduced count statistics.

\subsection*{Spectral Modelling}
\label{spectralmodelling}

We analyze all datasets in the SPEX fitting package \cite{Kaastra+96} using Cash statistics \cite{Cash+79}. The high-resolution \xmm\ RGS data might not have sufficient counts per bin in all of the analyzed spectral bins, thereby preventing us from applying the standard $\chi^2$ analysis instead. We adopt a distance of Her X-1 of 6.1 kpc \cite{Leahy+14}. All uncertainties are provided at \delcstat$=1$ significance.

The $0.3-10$ keV spectrum of Her X-1 is very complex, and requires many components to describe it accurately. We model the spectral continuum phenomenologically to decrease the computational cost, allowing us to apply a physical model for the ionized absorption from the disk wind. The broadband spectrum must be described well with the continuum model for us to correctly fit the absorption features, which appear very weak at later phases of the precession cycle.

The continuum model and its justification is described in much more detail in Paper I (section 4). In brief, we use a Comptonization \textsc{comt} model \cite{Titarchuk+94} to describe the primary accretion column radiation \cite{Lamb+73, Ghosh+79}, two blackbodies (\textsc{bb} and \textsc{mbb}) for the soft X-ray reprocessed emission \cite{Hickox+04}, three Gaussians for the Fe K energy band emission lines (Fe I narrow line, and two broad lines), a broad Gaussian for the Fe L emission around 1 keV \cite{Fuerst+13}, and four more Gaussians to describe the soft X-ray emission lines below 1 keV \cite{Kosec+20} (broad O VIII and N VII lines, narrow O VII and N VI intercombination lines). 

The disk wind absorption is modelled with a \textsc{pion} component \cite{Miller+15, Mehdipour+16}. \textsc{pion} determines the transmission of a slab of photo-ionized plasma by self-consistently calculating the absorption line opacities using the ionizing balance obtained from the spectral energy distribution (SED) of the current continuum model. By fitting the disk wind absorption lines with this model, we are able to recover the outflow column density, ionization parameter $\log \xi$, the systematic velocity of the outflow as well as its velocity width. The velocity width could be due to turbulent motions within the outflow, due to centrifugal flows or due to time variable systematic velocity.

Neutral ISM absorption is applied to the full spectral model via the \textsc{hot} component. The column density towards Her X-1 is very low \cite{HI4PI+16}, around $1\times10^{20}$ \pcm. Additionally, we found an issue with the gain of the EPIC pn instrument resulting in energy shifts of detected photons. The shifts are significant, of the order of 100 eV at $6-7$ keV and detrimental to our spectral analysis. This issue is described in much more detail in Appendix A of Paper I. It is corrected by applying a shift component \textsc{reds} to our final spectral model that effectively blueshifts the EPIC pn model by a certain amount. The blueshift is unknown a priori and is a fitted variable. For most observations, we find it to be in the range between $0.01-0.02$. Finally, we found strong residuals in the $22.25-24$ \AA\ wavelength range in the RGS spectra which might be of instrumental origin, and ignore this region in the present analysis. These residuals are discussed in Appendix B of Paper I. Ignoring them does not affect the present disk wind study as this wavelength band does not contain any relevant wind absorption lines.

\subsection*{The non-Solar elemental abundances of Her X-1}

Her X-1 is known to show strongly non-Solar elemental abundances, indicating enrichment of the secondary HZ Her envelope by the CNO process elements, leading to over-abundance of N and under-abundance of C \cite{Boroson+97, Jimenez+02}. The N/O ratio was found to be as high as 4, and Ne was also found to be over-abundant with respect to O by previous X-ray emission line studies \cite{Jimenez+02, Jimenez+05}. The N over-abundance is further supported by UV band studies \cite{Raymond+93}. Therefore, to describe the disk wind absorption lines accurately, we also need to fit for the elemental abundances. Our initial Her X-1 wind study \cite{Kosec+20} attempted to fit the abundance ratios by fixing one of the strong elements (O or Fe) to abundance of 1.0 and freeing the abundances of N, Ne and either Fe or O. We found an over-abundance of N/O by about $2-5$, an over-abundance of Ne/O of roughly $2-3$, and an Fe/O ratio as high as 10 (but poorly constrained) for the ionized wind gas.

Due to the computational expense required, we are unable to perform a similar analysis for all 28 analyzed spectra simultaneously. Since we do not expect the abundances to vary between the individual observations or segments, we choose to fit the elemental abundances only for one observation, and fix these best-fitting abundances for all other analyzed segments. We chose the full high flux part of observation 0865440101 (about 50 ks of clean exposure, split into segments 2 to 6 in the final time-resolved analysis), which showed the strongest absorption lines from the wind. Since all of the wind absorption signatures at these ionization parameters (\logxi$\sim3.5-4$) are metallic lines (H is practically fully ionized), we had to freeze the abundance of one of the elements and measure the remaining elemental abundances with respect to it. We fixed the O abundance to 1 as the O VIII line is the strongest absorption line in the well-resolved RGS energy band. The abundances of N, Ne and Fe were freed. The absorption lines of the remaining elements were too weak to provide meaningful constraints.

We find the following elemental abundances (with respect to the fixed O abundance): (N/O)$=3.4_{-0.8}^{+0.6}$, (Ne/O)$=2.3_{-0.5}^{+0.4}$ and (Fe/O)$=2.1 \pm 0.3$, consistent with our previous findings except for the now much more realistic Fe/O ratio. The difference in the (Fe/O) ratio comparing with our archival Her X-1 study is likely due to our much better spectral model (see the following section for further details) as well as the much better spectral statistics. The recovered best-fitting elemental abundances from this observational segment are fixed in all the subsequent time-resolved spectral fits. From these spectra we also obtain a disk wind column density of $1.00^{+0.09}_{-0.05} \times 10^{23}$ \pcm, an ionization parameter \logxi\ of $3.77_{-0.02}^{+0.03}$, a velocity width of $186 \pm 15$ km/s, and a systematic velocity of $-420 \pm 40$ km/s. The spectral fit is shown in \ref{101_spectrum}. 

We note that the very low uncertainties on the ionization parameter are due to very high EPIC pn statistics in the Fe K band (in this stacked observation), strongly constraining the ratio of the Fe XXV and Fe XXVI lines, which drives the value of $\log \xi$. The shorter observations and observation segments have poorer statistics, generally resulting in larger uncertainties on the ionization parameter (\ref{resulttable}). The individual segments show a range of Fe XXV and XXVI line ratios, from spectra where the Fe XXV line optical depth exceeds the Fe XXVI depth (\logxi $\sim 3.0$), to spectra where Fe XXVI dominates (\logxi $\sim 3.9$), providing a model-independent check of our photoionization absorption spectral models.

The O abundance might not in reality be equal to 1 as fixed in our spectral fit, which might have an effect on our wind parameter measurements. Unfortunately, it is impossible to constrain an absolute elemental abundance using our photoionized absorber fits. To test how much this assumption affects our measurements, we fit observation 0865440101 by the same spectral model as above but instead of fixing O abundance to 1, we fix it to 0.7 and 1.5, respectively. We do not expect the O abundance to be wildly different from 1, as the CNO cycle will create a strong over-abundance of N and a strong under-abundance of C, without heavily depleting the O abundance \cite{Przybilla+10}. Additionally, no O depletion was observed in Her X-1 by the studies of its narrow X-ray emission lines \cite{Jimenez+02}. 

As expected, the best-fitting abundances of N, Ne and Fe shift accordingly to conserve broadly consistent abundance ratios as obtained in the original fit, and the inferred outflow velocities and velocity widths are not strongly affected. The ionization parameter is not affected by the choice of absolute O abundance either, as it is constrained by the ratio of Fe XXV and XXVI lines. Finally, we do observe differences in the column density of about $10-20$\% compared with the original fit. Importantly, this error would occur (in the same direction) for all analyzed observations as the O abundance does not to vary in time, systematically shifting our column density measurements lower or higher, depending on the actual absolute O abundance. As a result, the measured distances from the X-ray source would (linearly) suffer from the same systematic, and so would the calculated absorber heights above the disk. Therefore, importantly, this possible systematic error does not change our qualitative results about the weakening and fragmenting outflow which expands to larger heights and distances from the neutron star. In general, our statistical uncertainties from the individual segment measurements are comparable or larger than this potential systematic.

\subsection*{Comparison with our pilot study of Her X-1}

In agreement with our first wind study of Her X-1 \cite{Kosec+20}, the optical depth of the ionized absorption features decreases strongly with the increasing precession phase. In addition to the visual confirmation in \ref{xcamp_opt_depths}, the decrease in line optical depths can be seen through the significant reduction in \delcstat\ fit improvements upon adding the ionized absorber to the broadband continuum model (\ref{delcstat_vs_suporb}). While the lines are very strong at early precession phases and adding the wind absorption model often improves the fit by \delcstat$>100$ (statistical significance of outflow detection $>>5\sigma$), at later phases the statistical change is much lower, with \delcstat$\sim25$ or less. We note that the archival \xmm\ and \chandra\ observation exposure times (and data quality) are not correlated with precession phase, and the August 2020 observational segment exposure times increase with precession phase. Nevertheless, the fit improvement dramatically decreases with increasing phase, indicating that the absorption lines are disappearing.

The strong optical depth variation of the absorption lines can be explained by either a decrease in the column density of the ionized absorber, or an increase in its ionization parameter (or some combination of both variations). In our first paper using the archival \xmm\ datasets \cite{Kosec+20}, we concluded that the variation was driven by an increase in the ionization parameter of about 2 orders of magnitude over the sampled range of precession phases. Now, with our significantly improved continuum model, a much higher quality dataset, and with better treatment of the EPIC pn gain than in the original study, we have a higher fidelity photoionization model. We find that the ionization parameter does not increase with precession phase (instead it decreases), but the optical depth changes are primarily driven by a very strong decrease of the wind column density with precession phase ($10^{23}$ to $10^{20}$ \pcm). Our original conclusion about the parameter variation was likely caused by the simplistic modelling of the Fe K band (poorly resolved by EPIC pn), where the Fe emission was only described with a single broad Gaussian, as well as due to not including a correction factor to the EPIC pn gain. In this paper, we show that with these improvements in analysis, the re-analyzed archival \xmm\ data exhibit the same trend in column density and ionization parameter as the new XMM-Newton and Chandra data.

We still observe some scatter between the neighbouring wind measurements. This could be due to physical changes of parameters in a clumpy wind, but could alternatively be due to our limited spectral resolution with \xmm\ in the Fe K band. To some extent EPIC-pn has difficulties separating the three emission components of Fe of various line widths (Paper I) from the two narrow wind absorption lines (Fe XXV and XXVI). The ratio of the two Fe absorption lines is important in constraining the ionization parameter (and the column density) of the wind. \chandra\ offers better spectral resolution at these energies but much poorer statistics due to a lower effective area. Her X-1 will thus be a prime target for the forthcoming \xrism\ observatory \cite{xrism+20}, to be launched in 2023. Thanks to a combination of excellent spectral resolution and high collecting area, \xrism\ will be able to \textit{precisely} measure the physical conditions at various heights of the outflow in Her X-1. Finally, some of the observed scatter in the archival \xmm\ and \chandra\ observations (which occurred many precession cycles before the August 2020 campaign) could potentially be attributed to long-term changes in the wind structure. We particularly note that all three strongest outliers in Fig. \ref{constdist_location} correspond to archival \xmm\ measurements.

\subsection*{Comparison of a simultaneous \xmm\ and \chandra\ observation}

Two of the \chandra\ observation occurred in August 2020, during the intensive \xmm\ coverage. One of these observations happened simultaneously with the third \xmm\ exposure, at the precession phase of $\sim0.125$. We can therefore compare the best-fitting wind parameters from the \xmm\ and \chandra\ instruments. We find that the measured wind column density, outflow velocity and turbulent velocity agree within 90\% uncertainties between the two instruments. We do find a difference between the best-fitting ionization parameters, of about \logxi $\sim0.3$. This is larger than the measured uncertainties, possibly indicating some systematic difference between \chandra\ and \xmm, at least in August 2020. We note that the August 2020 campaign occurred after \chandra\ experienced significant degradation of the soft energy effective area, particularly below 1 keV. During this observation the counts in the soft band are so low that the important O VIII-N VII region has poor signal-to-noise. With this energy band unusable, it is therefore possible that the \chandra\ measurement underestimates the ionization parameter of the outflow.

\subsection*{Potential degeneracies in the disk wind spectral modelling}

The spectrum of Her X-1 is highly complex in the $0.3-10$ keV energy band. As a result, our spectral model has 33 free parameters (emission components + disk wind), all listed in Table 4 of Paper I. It is possible that the resulting fitting freedom could introduce degeneracies between the disk wind and emission component parameters. Below we explore these potential issues, to assess how robust are our disk wind measurements.

We used the Bayesian inference tool Multinest \cite{Feroz+09} to explore the parameter space which may contain multiple posterior modes and degeneracies between model parameters. We follow the Bayesian analysis method developed for the SPEX fiting package \cite{Rogantini+21} based on the Bayesian X-ray Analysis tool \cite{Buchner+14}. Due to the required computational expense, we are unable to apply the nested sampling algorithm to the full observational sample, or consider all the 33 free parameters at once. However, most of the emission component parameters are unlikely to influence the disk wind fitting. Any broadband component parameters are constrained by the continuum away from the narrow disk wind absorption lines. This is also the case for the broadened and well-resolved emission lines within the RGS band (Fe L, O VIII and N VII). Finally, the narrow emission lines in the RGS band (the intercombination lines of O VII and N VI) are located away from any strong wind absorption.

The emission component parameters which are more likely to be degenerate with the disk wind absorption are the properties of the 3 emission lines in the Fe K band, which is only moderately resolved with EPIC pn. Additionally, it is unclear what effect the gain shift parameter could have on the disk wind properties. For this reason, we perform the nested sampling study over the following 11 parameters: 6 free parameters of the Fe K band lines, 1 gain shift parameter and 4 disk wind parameters. In \ref{MCMCfig}, we show the results as a corner plot for the high flux part of observation 0865440101 (segments 2-6 stacked). The disk wind parameters are strongly constrained and we do not find any significant unexpected correlations between the wind parameters themselves, or between the disk wind and the continuum parameters. We observe a mild correlation between the column density and the ionization parameter, but this is expected in photo-ionization plasma modelling. There is also a mild correlation between the column density and the gain shift, but the gain shift is still strongly constrained at better than 10\% level (at 1$\sigma$). Finally, we do find correlation between some parameters of the two broadened Fe emission lines (medium-width Fe XXV and broad Fe K), but these do not appear to cause any issues with the wind parameter recovery and the emission line parameters are still fairly well constrained (10\% at 1$\sigma$ or better).

We also performed the same nested sampling analysis on 4 individual observation segments, representatively selected to evenly sample the observed range of disk precession phases. These MultiNest outputs show comparable results to the first one. No obvious significant correlations are found between the emission line parameters or gain shift and the disk wind parameters, except for the expected correlation between the wind column density and its ionization parameter.

Finally, we also consider the possibility that the disk wind in Her X-1 is multi-phase. In such a case, modelling its absorption signatures with a single photoionized component could result in systematic errors in the recovered best-fitting parameters. We previously considered this hypothesis for the archival Her X-1 \xmm\ datasets \cite{Kosec+20}, but found no significant evidence for multi-phase nature of the outflow.

We perform the same test for new datasets from August 2020. We take the best-fitting spectral models for all individual segments from 2020, add a second photoionized absorber (a second \textsc{pion} component) to the original fit, and perform the fit and the error search again. We explored a range of different \logxi\ parameters of the potential second wind phase, between \logxi\ of 1.5 and 4.5. No significant detections are found in any of the observational segments. In all cases the fit improvement over the baseline single-phase outflow model is \delcstat$<10$, indicating weak or no evidence for the multi-phase outflow hypothesis.

\subsection*{The wind mass outflow rate}

The mass outflow rate of a disk wind can be determined using the measured wind ionization parameter $\xi$, outflow velocity $v$ and the ionizing luminosity $L_{ion}$ of the X-ray source, by applying Eq. 7 derived in our pilot Her X-1 study \cite{Kosec+20}:

\begin{equation}
\label{eqfinalMout}
\dot{M}_{out} = 4 \pi \mu m_{p} v \frac{L_{ion}}{\xi}  C_{V}  \frac{\Omega}{4 \pi}
\end{equation}

where $C_{V}$ is the volume filling factor, which is likely high (approaching unity) under our first assumption of $\Delta R/R \sim1$ for the relative thickness of the absorbing layer. Under the second assumption of constant wind density, $C_V$ must decrease at larger distances from the neutron star (since $\Delta R/R$ decreases). $\Omega$ is the solid angle into which the outflow is being launched, $\mu$ defines the mean atomic mass ($\sim1.2$ assuming solar abundances) and $m_p$ is the proton mass. We note that there is an alternative way to estimate the mass outflow rate by assuming that the observed wind velocity is equal to the local escape velocity ($v^{2} \approx 2GM/R$, where $M$ is the compact object mass). However, this assumption is unlikely to be applicable in the case of Her X-1 considering we observe no correlation between the outflow velocity and any other relevant parameters of the system. 

We make a rough estimate of the wind mass outflow rate by setting both $\Omega/4 \pi$ and  $C_V$ to unity (thus following the first assumption about the large relative thickness of the wind layer). The results are shown in \ref{absdist_mass_out}, in addition to the mass accretion rate through the outer accretion disk, which is roughly $7\times10^{-9}$ M$_{\odot}$ year$^{-1}$, measured via UV observations \cite{Boroson+07}. The mass outflow rates are about 10 times the mass accretion rate at low precession phases (with the exception of phase$\sim0$), indicating that the wind launch solid angle is likely significantly lower than $4\pi$, in agreement with the outflow being an equatorial wind with a small opening angle. Reasonable mass outflow rates can be recovered by setting $\Omega/4\pi \sim0.1$ (opening angle of the wind of about 5-10$^{\circ}$). Hence a significant fraction of the originally accreting matter could be launched into an ionized outflow.

However, the estimated mass outflow rates significantly increase with the precession phase, to at least $\sim100$ times the mass accretion rate (without the $\Omega$ correction factor) towards the end of the \textit{Main High State}. The mass outflow rate should not be greater than the accretion rate at any point, considering the long-term stable accretion in Her X-1. This indicates that the volume filling factor $C_{V}$, and thus the relative thickness $\Delta R/R$ must be much lower than 1 at later phases (at larger distances from the neutron star) and be a decreasing variable with phase/distance from the X-ray source. These findings further motivate our second assumption of constant density with distance - a fragmenting/clumping outflow rather than a tenuous, high volume filling but very low density disk wind.

\subsection*{Modelling the warped disk shape}

We adopt a warped disk model based on the Her X-1 precession cycle and X-ray pulse profile evolution \cite{Scott+00}. This allows us to make a first attempt to describe the vertical structure of the disk wind in Her X-1. The disk is modelled as an array of tilted rings with radii varying between the inner and outer disk radius, with a twist between the consecutive rings. The assumed tilt of the outer accretion disk is 20$^{\circ}$, the tilt of the inner accretion disk is 11$^{\circ}$, with a twist between the inner and the outer radii of 138.6$^{\circ}$ (the inner radius leading the precession). These angles were determined from the transition times between the \textit{High} and the \textit{Low States}\cite{Scott+00}. The evolution of the tilt and twist with radius within the disk is not given and depends on the physical mechanism introducing and maintaining the disk warp \cite{Schandl+94, Ogilvie+01}. As a first order approximation, we assume linear evolution of both quantities. We use an outer disk radius of $2\times10^{11}$ cm \cite{Cheng+95}, and an inner radius of $2\times10^{8}$ cm (calculated by assuming a dipolar Her X-1 magnetic field). We note that our choice of the inner disk radius introduces a negligible error in our wind position calculation since we assumed linear evolution of the tilt and the twist angles with radius (and the disk spans multiple orders of magnitude in size).

The radii within the disk are obtained from the absorber distances from the X-ray source using the assumption of constant absorber density (bottom panel of Fig. \ref{abs_dist}). This is likely a more realistic estimate of the outflow position than the one obtained assuming constant relative absorber thickness (top panel of Fig. \ref{abs_dist}), which results in an unrealistically large precessing wind structure and extreme mass outflow rates. Finally, we assume a Her X-1 binary plane inclination of 85$^{\circ}$, i.e. an almost edge-on disk \cite{Leahy+14}. By applying all of the above, we obtain the vertical position at which our sightline crosses the outflow structure, for all individual observations or observation segments. Combining the measurements of the height above the disk and the distance from the X-ray source, we create the 2D color map of the accretion disk wind properties (Fig. \ref{constdist_location}). The calculated heights are also listed in \ref{resulttable_calculations}.

 \newpage

\begin{figure*}
\begin{center}
\includegraphics[width=0.9\textwidth]{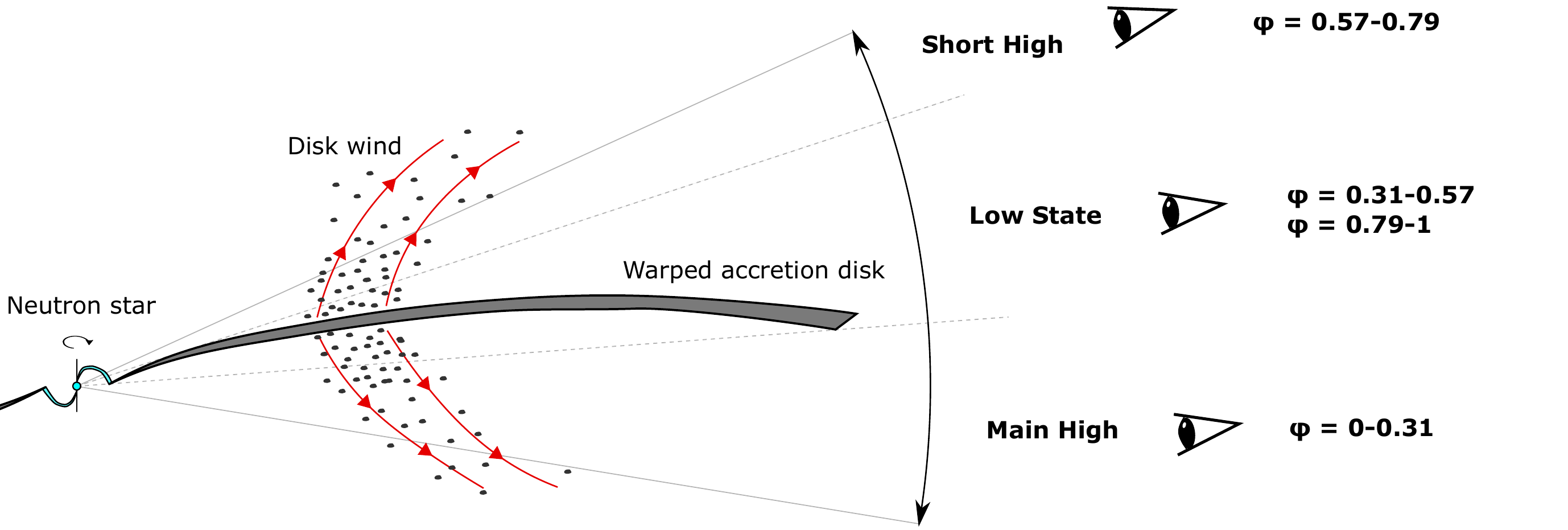}
\caption{\textbf{A schematic of the Her X-1 precession cycle.} A simplified schematic of the precession cycle (not to scale), where the disk is stationary and our line of sight relative to the disk is in motion, showing the different states of the 35-day cycle along with the appropriate precession phases $\phi$. Our \xmm\ and \chandra\ observations sample the range of precession phases from $\phi=0$ to $\phi\sim0.2$. During the \textit{Low States}, the disk obscures our line of sight towards the neutron star, while during the \textit{High States}, the neutron star is directly visible. The observed X-ray flux changes are thus an obscuration effect while the long-term mass accretion rate is roughly constant, as shown by the stable X-ray irradiation of the secondary HZ Her by Her X-1 \cite{Gerend+76}. As the sightline towards the X-ray source changes during the \textit{High States}, different parts of the wind structure (at various heights above the disk) are sampled, allowing us to uniquely measure the variation of wind properties along the streamlines. In reality, the disk shape also varies over the precession cycle, which cannot be shown in this simplified schematic. \label{HerX1_scheme}}
\end{center}
\end{figure*}

\begin{figure*}
\begin{center}
\includegraphics[width=0.65\textwidth]{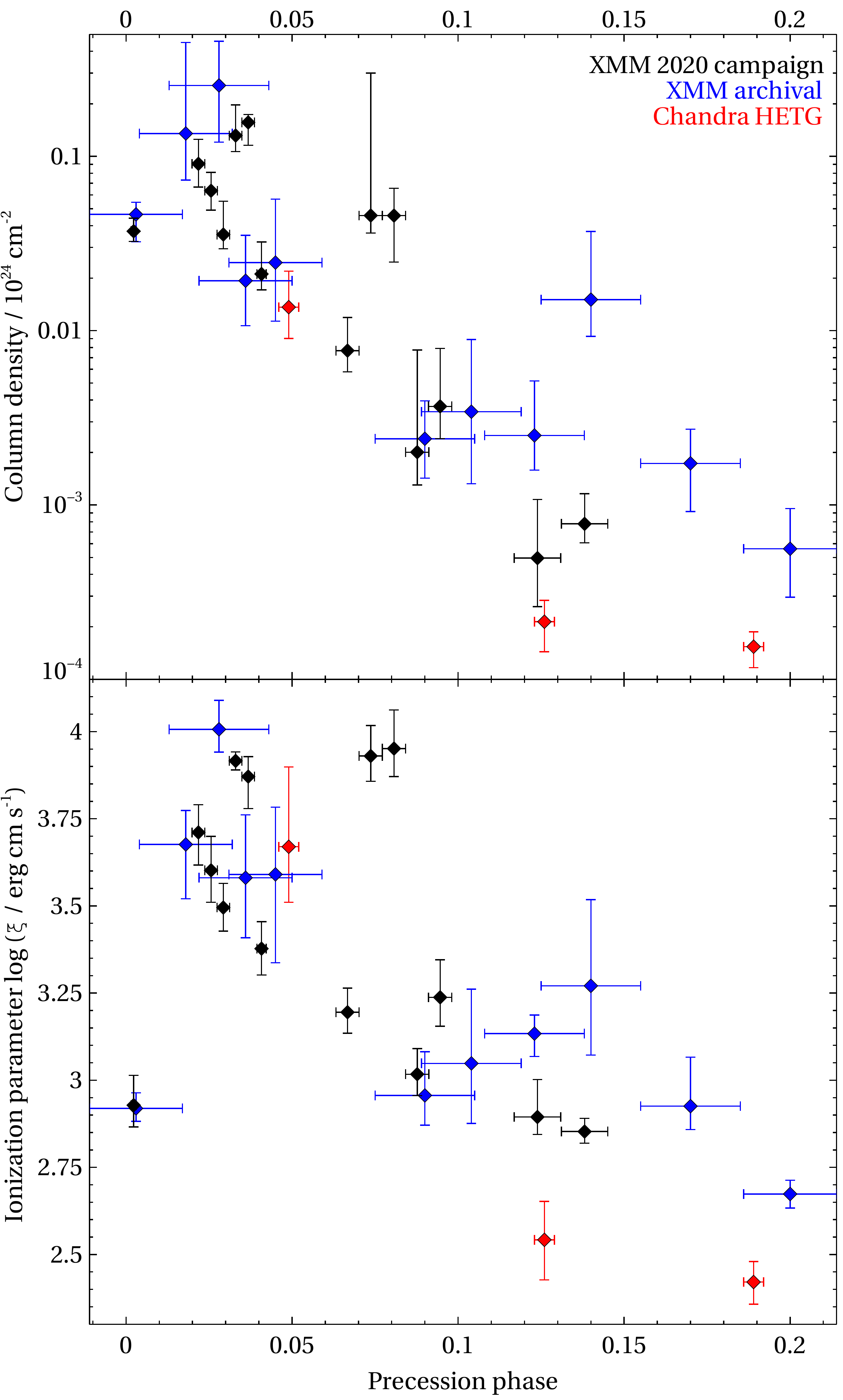}
\caption{\textbf{Evolution of the wind column density and ionization parameter with precession phase.} Variation of the disk wind column density (top panel) and ionization parameter (bottom panel) versus the disk precession phase. Observations from the August 2020 \xmm\ campaign are in black, archival \xmm\ observations are in blue, and \chandra\ observations are in red color. \label{nh_xi_suporb}}
\end{center}
\end{figure*}

\begin{figure*}
\begin{center}
\includegraphics[width=0.65\textwidth]{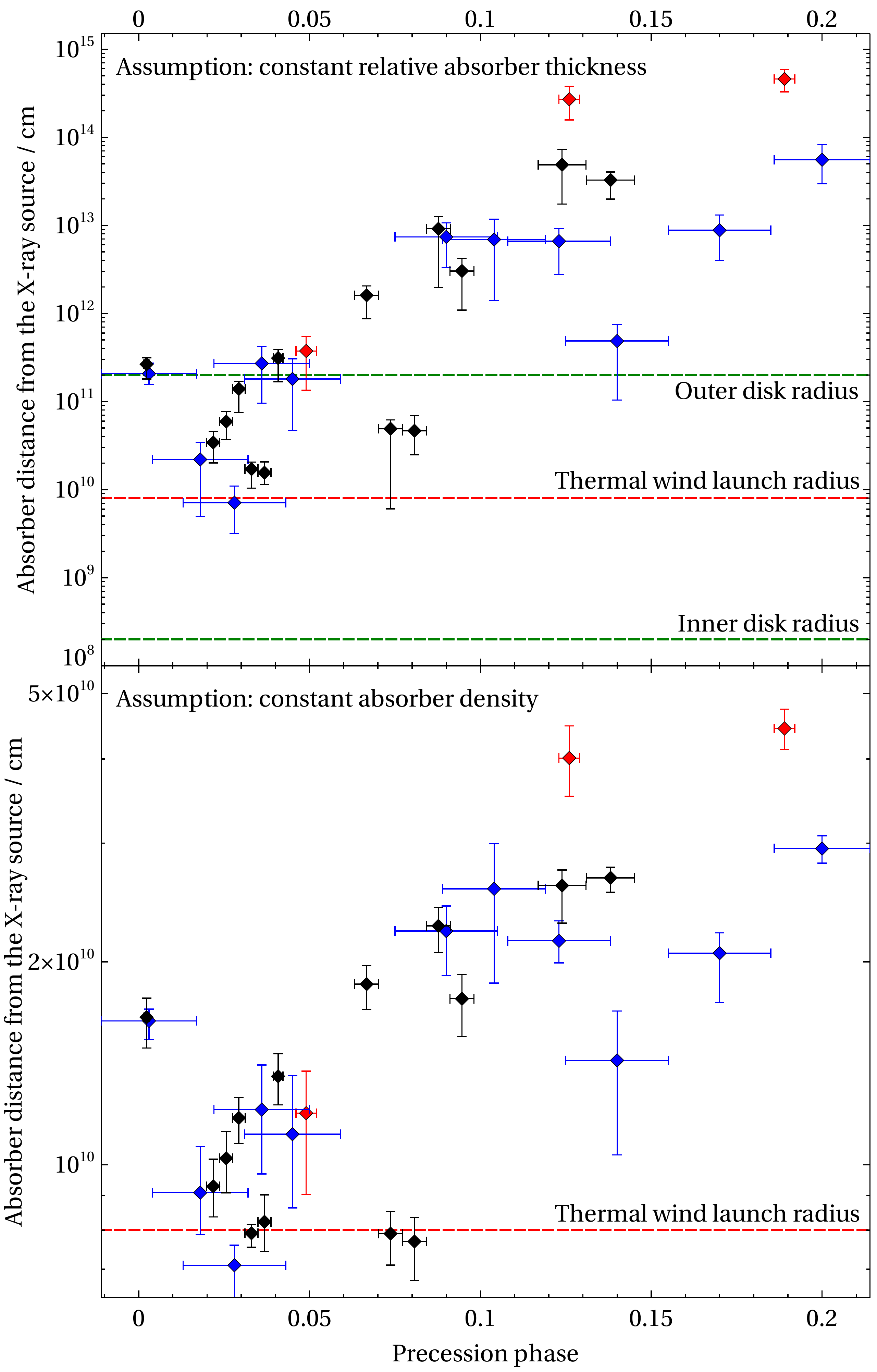}
\caption{\textbf{Measured distances of the outflow from the X-ray source.} Top panel: The ionized wind distance from the X-ray source, assuming its relative thickness ($\Delta R/R$) is constant and equal to unity. Bottom panel: The wind distance from the X-ray source, assuming its density is constant at all locations, and equal to the density at the lowest distance data point in the top panel. In both panels, the August 2020 and archival \xmm\ observations are in black and blue, respectively, and \chandra\ observations are in red color. Horizontal dashed lines show the position of the inner, outer disk radii, as well as the calculated thermal wind launch radius \cite{Begelman+83, Kosec+20}. The calculated distances are listed in \ref{resulttable_calculations}. \label{abs_dist}}
\end{center}
\end{figure*}

\begin{figure*}
\includegraphics[width=\textwidth]{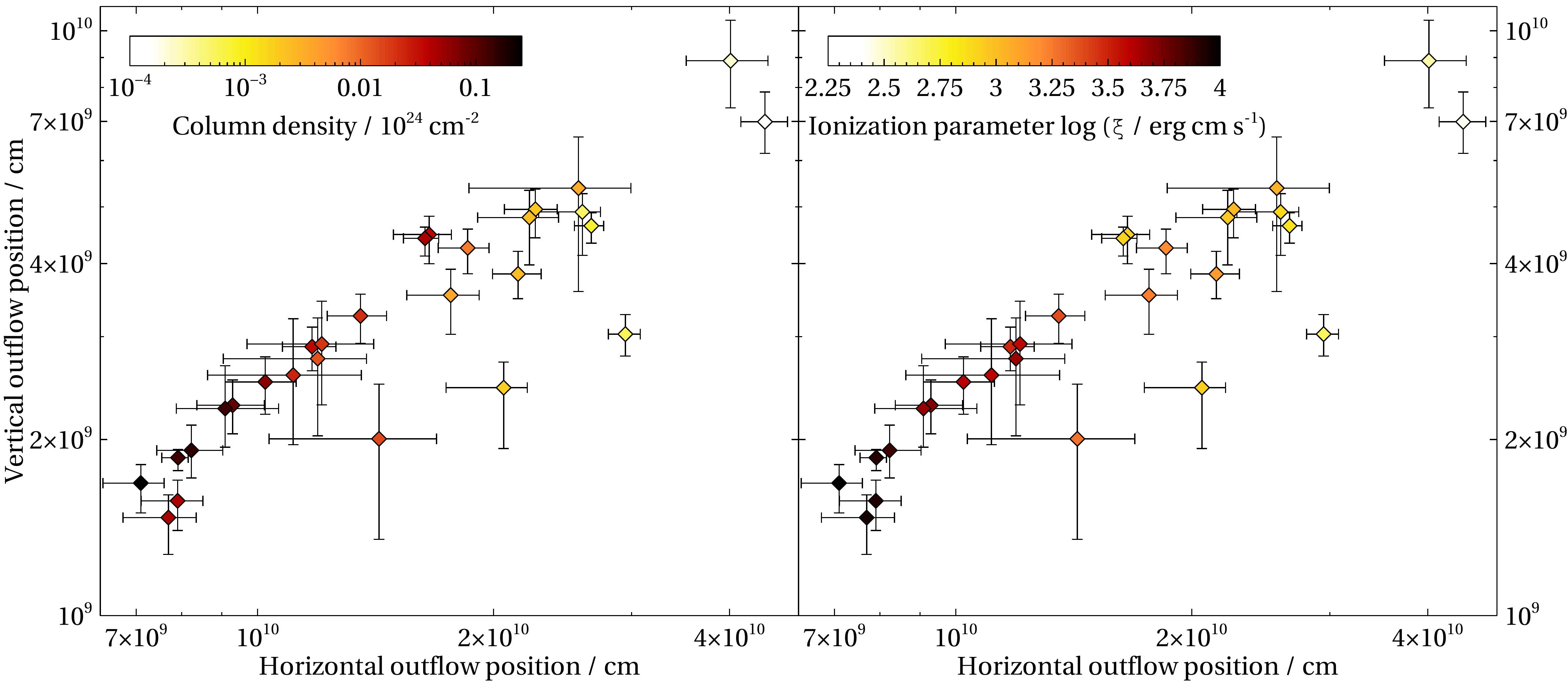}
\caption{\textbf{2D map of the disk wind properties.} Horizontal and vertical location of the ionized outflow during the 28 \xmm\ and \chandra\ observations was obtained by modelling the shape of the warped accretion disk. The color scale shows the measured outflow properties: column density is in the left panel, the ionization parameter is in the right panel. \label{constdist_location}}
\end{figure*}

\renewcommand\thefigure{Extended Data Figure \arabic{figure}}  
\renewcommand{\figurename}{}
\renewcommand\thetable{Extended Data Table \arabic{table}}  
\renewcommand{\tablename}{}
\setcounter{figure}{0}

\begin{figure*}
\includegraphics[width=\textwidth]{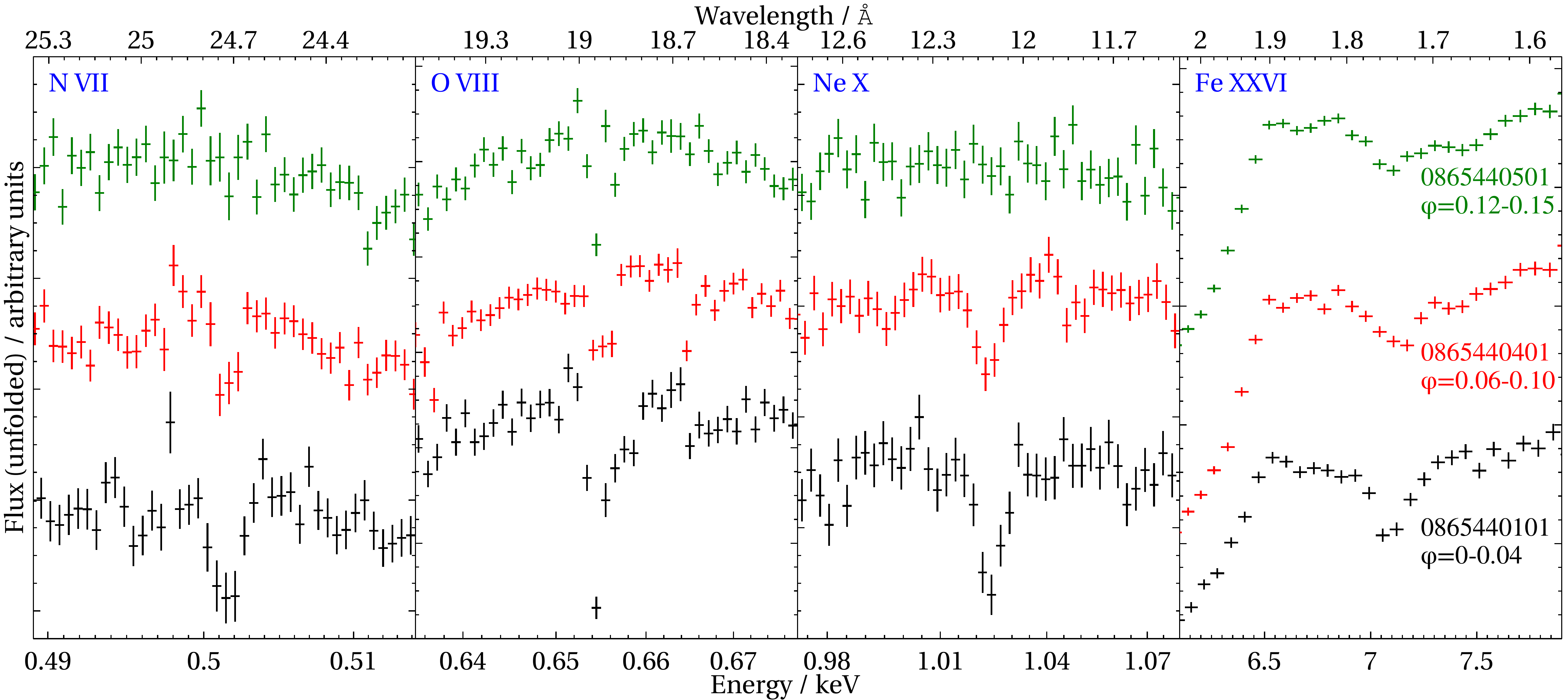}
\caption{\textbf{Optical depth of disk wind absorption lines with progressing precession phase.} Variation in optical depths of the disk wind absorption lines during the three full-orbit, back-to-back \xmm\ observations from August 2020. The 4 panels focus on the strongest elemental transitions of N VII, O VIII, Ne X and Fe XXV/XXVI. The Y-axis is in the units of Counts/cm$^{2}$/s/\AA, but the data from different observations were shifted vertically by linear amounts for visual purposes. Observation 0865440101, taken at the beginning of the precession cycle (phase 0-0.04), shows the strongest wind absorption. The lines get weaker at later precession phases during the second observation (0865440401, phase 0.06-0.10) and nearly disappear by the end of the third observation (0865440501, phase 0.12-0.15). \label{xcamp_opt_depths}}
\end{figure*}

\begin{figure*}
\includegraphics[width=\textwidth]{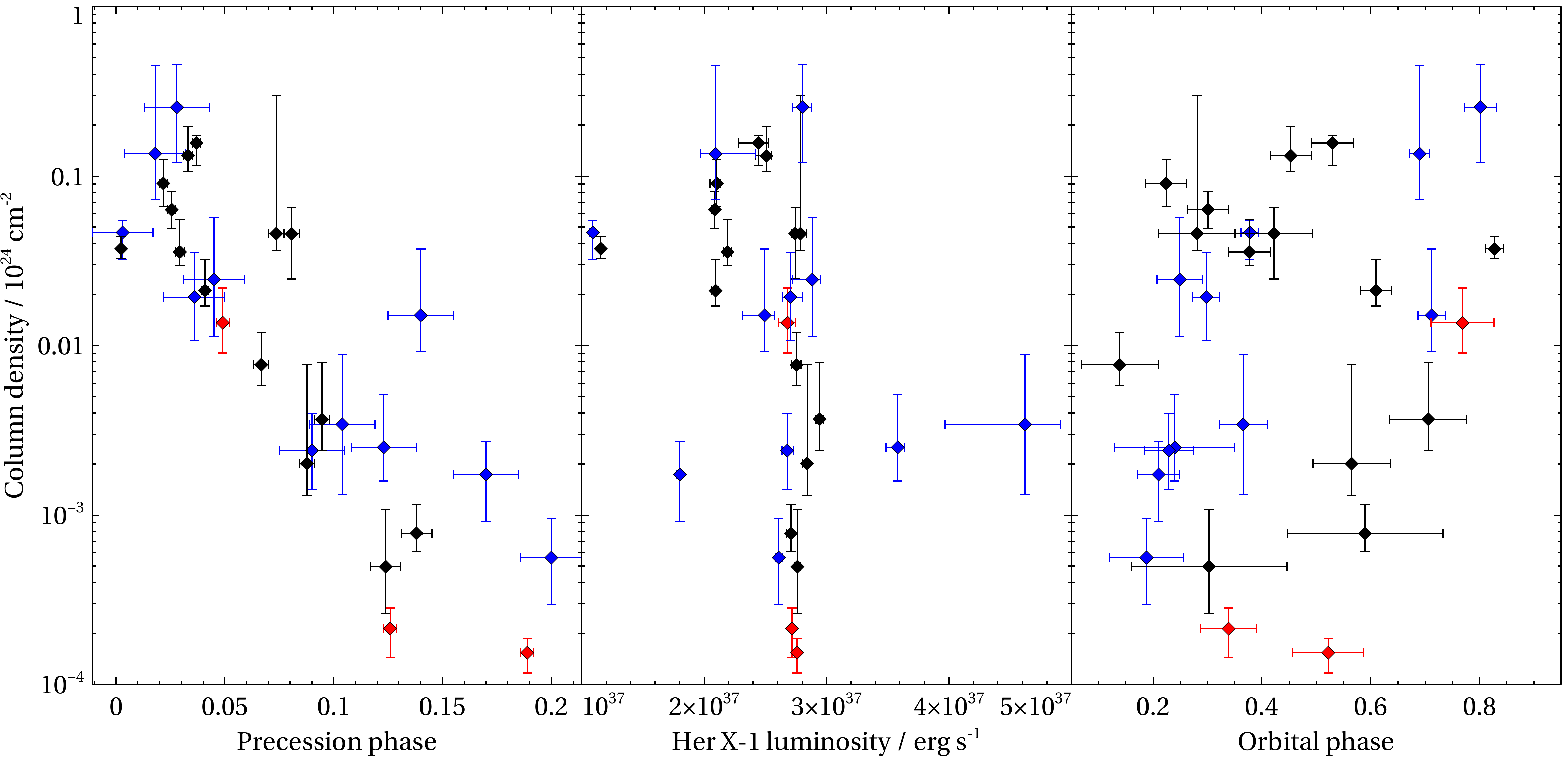}
\caption{\textbf{Evolution of the wind column density.} Variation of wind column density with precession phase (left panel), observed Her X-1 luminosity (middle panel) and orbital phase (right panel). Observations from the August 2020 and archival \xmm\ observations are in black and blue colors, \chandra\ observations are in red. \label{Evolution_nh}}
\end{figure*}

\begin{figure*}
\includegraphics[width=\textwidth]{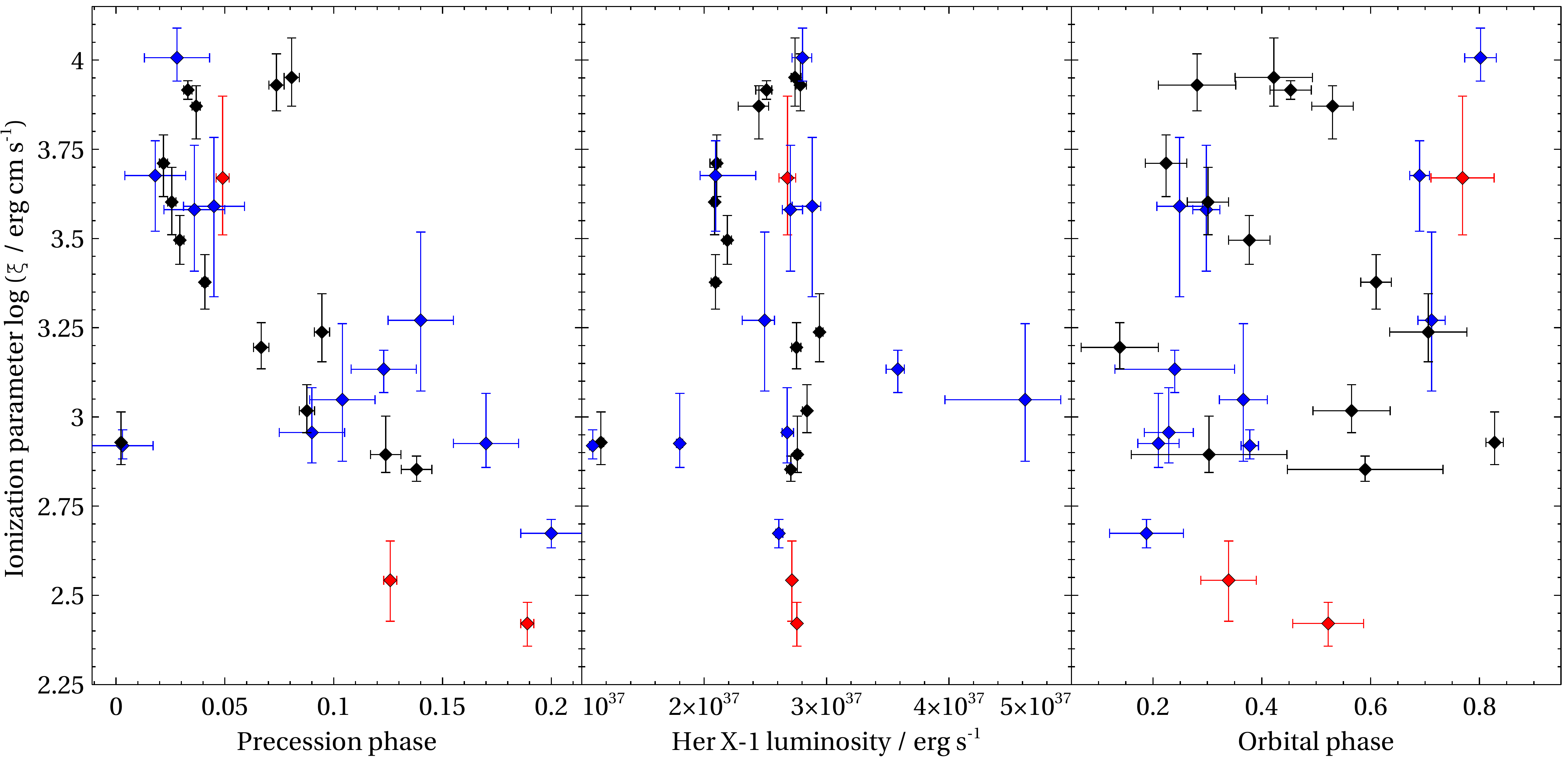}
\caption{\textbf{Evolution of the wind ionization parameter.} Variation of wind ionization parameter with precession phase (left panel), observed Her X-1 luminosity (middle panel) and orbital phase (right panel). Observations from the August 2020 and archival \xmm\ observations are in black and blue colors, \chandra\ observations are in red. \label{Evolution_xi}}
\end{figure*}

\begin{figure*}
\begin{center}
\includegraphics[width=0.75\textwidth]{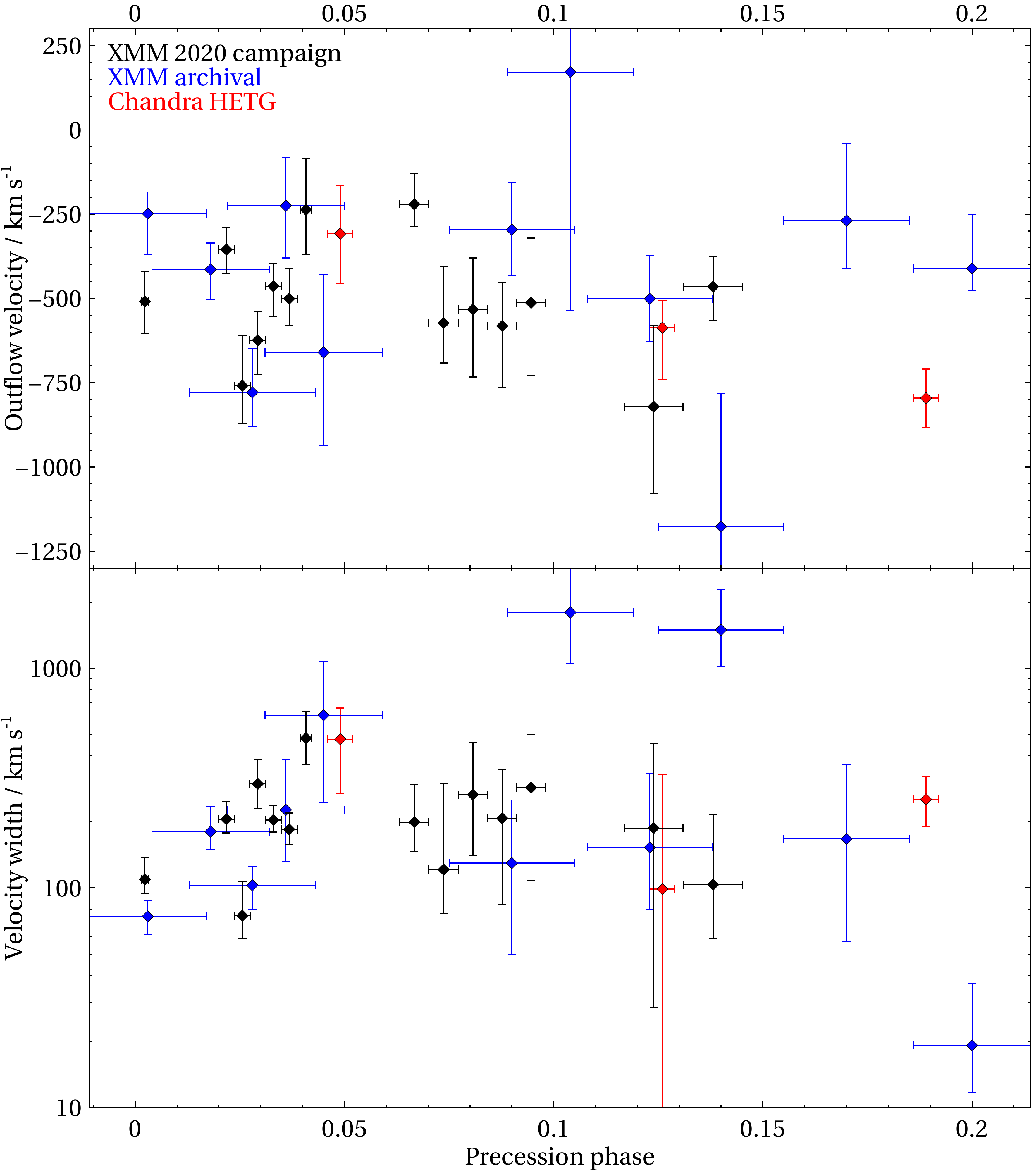}
\caption{\textbf{Evolution of the outflow velocity and velocity width with precession phase.} Variation of the disk wind outflow velocity (top panel) and velocity width (bottom panel) versus the disk precession phase. Observations from the August 2020 \xmm\ campaign are in black, archival \xmm\ observations are in blue, and \chandra\ observations are in red color. \label{zvcorr_vturb_suporb}}
\end{center}
\end{figure*}

\begin{figure}
\begin{center}
\includegraphics[width=\textwidth]{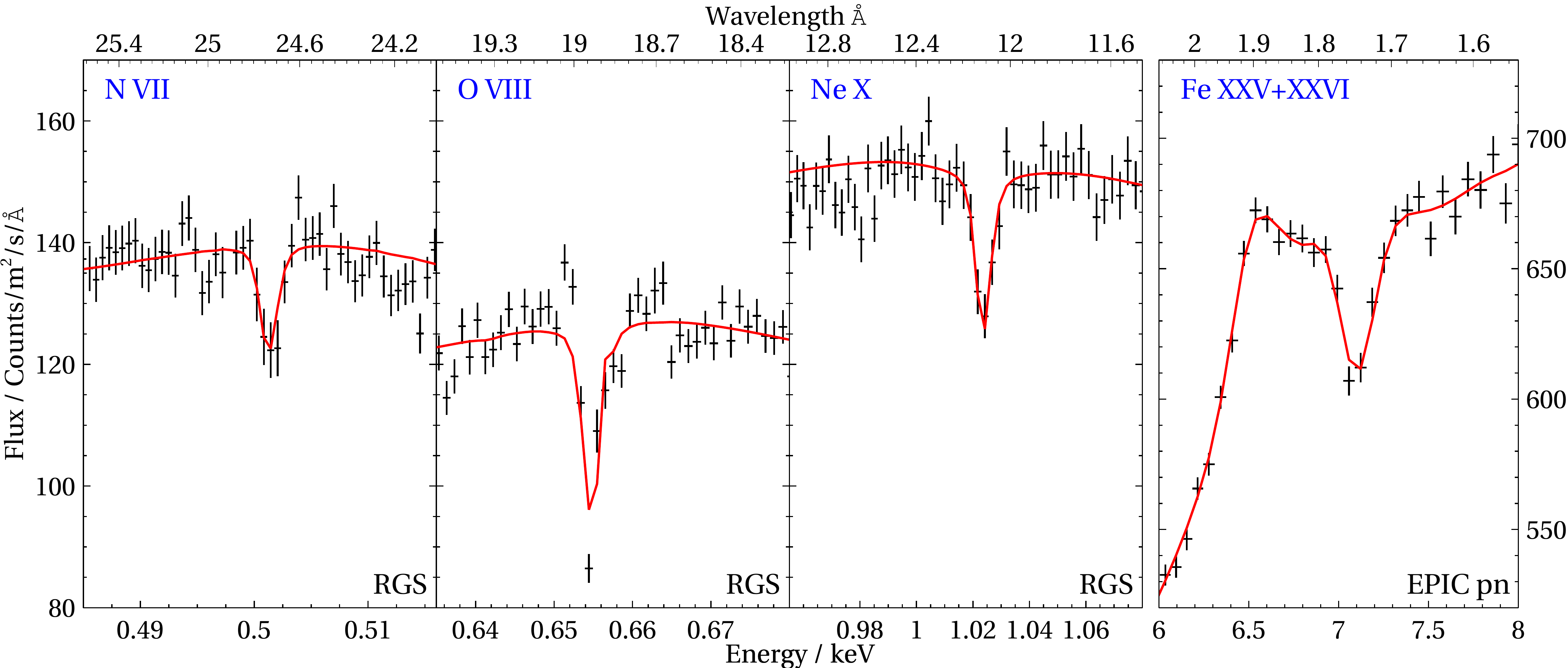}
\caption{\textbf{\xmm\ spectra from observation 0865440101.} RGS (first 3 subpanels) and EPIC pn (right subpanel) spectra from observation 0865440101 (segments $2-6$ stacked), fitted with the photoionized absorber model, focusing on the narrow energy bands around the strongest disk wind absorption lines of N VII, O VIII, Ne X and Fe XXV/XXVI. RGS 1 and 2 data are stacked for visual clarity. \label{101_spectrum}}
\end{center}
\end{figure}

\begin{figure}
\begin{center}
\includegraphics[width=0.6\textwidth]{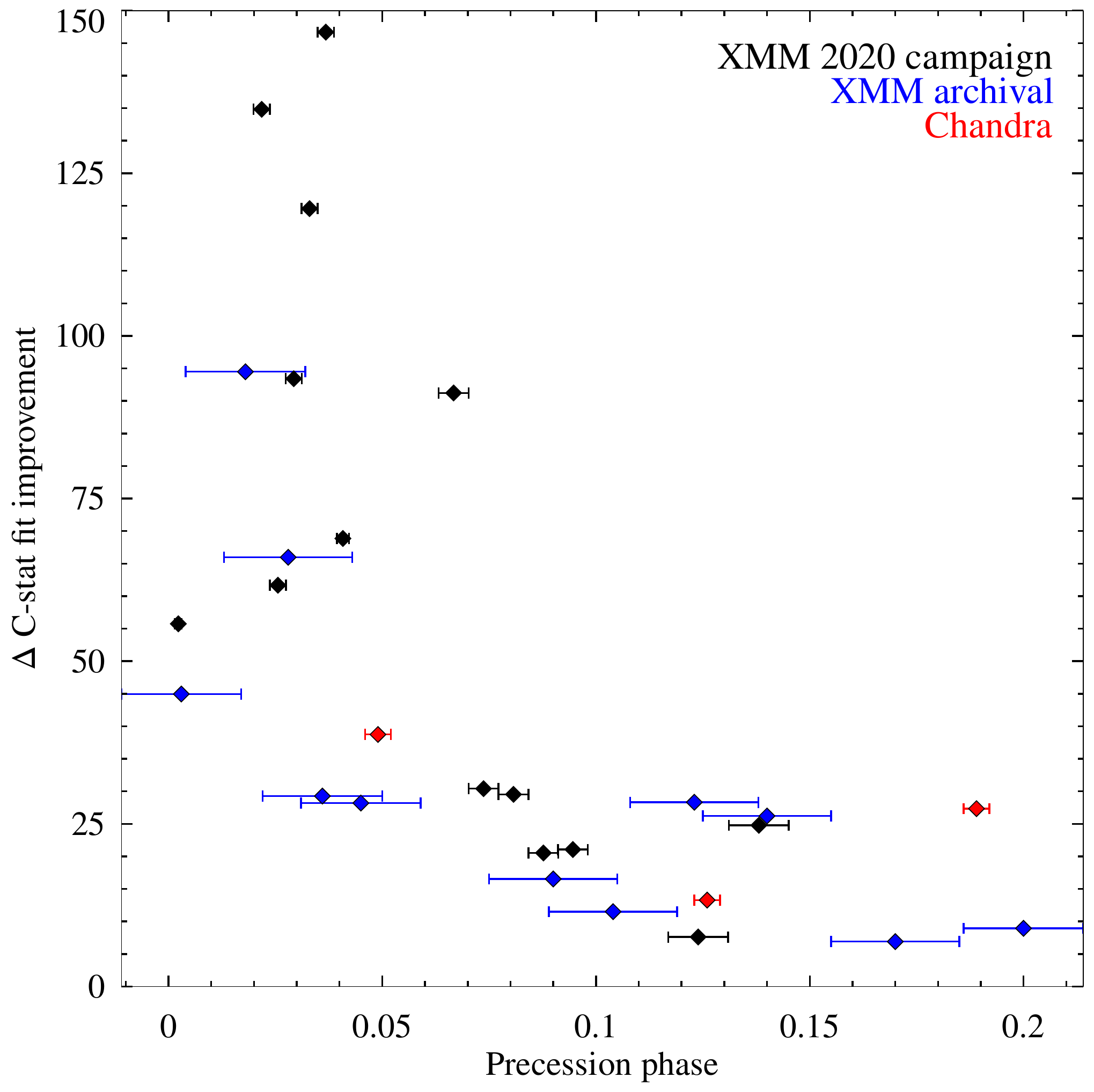}
\caption{\textbf{Statistical significance of outflow detection in each observation.} The \delcstat\ fit statistics improvement upon adding the ionized absorber to the broadband continuum spectral fit. \label{delcstat_vs_suporb}}
\end{center}
\end{figure}

\begin{figure}
\begin{center}
\includegraphics[width=\textwidth]{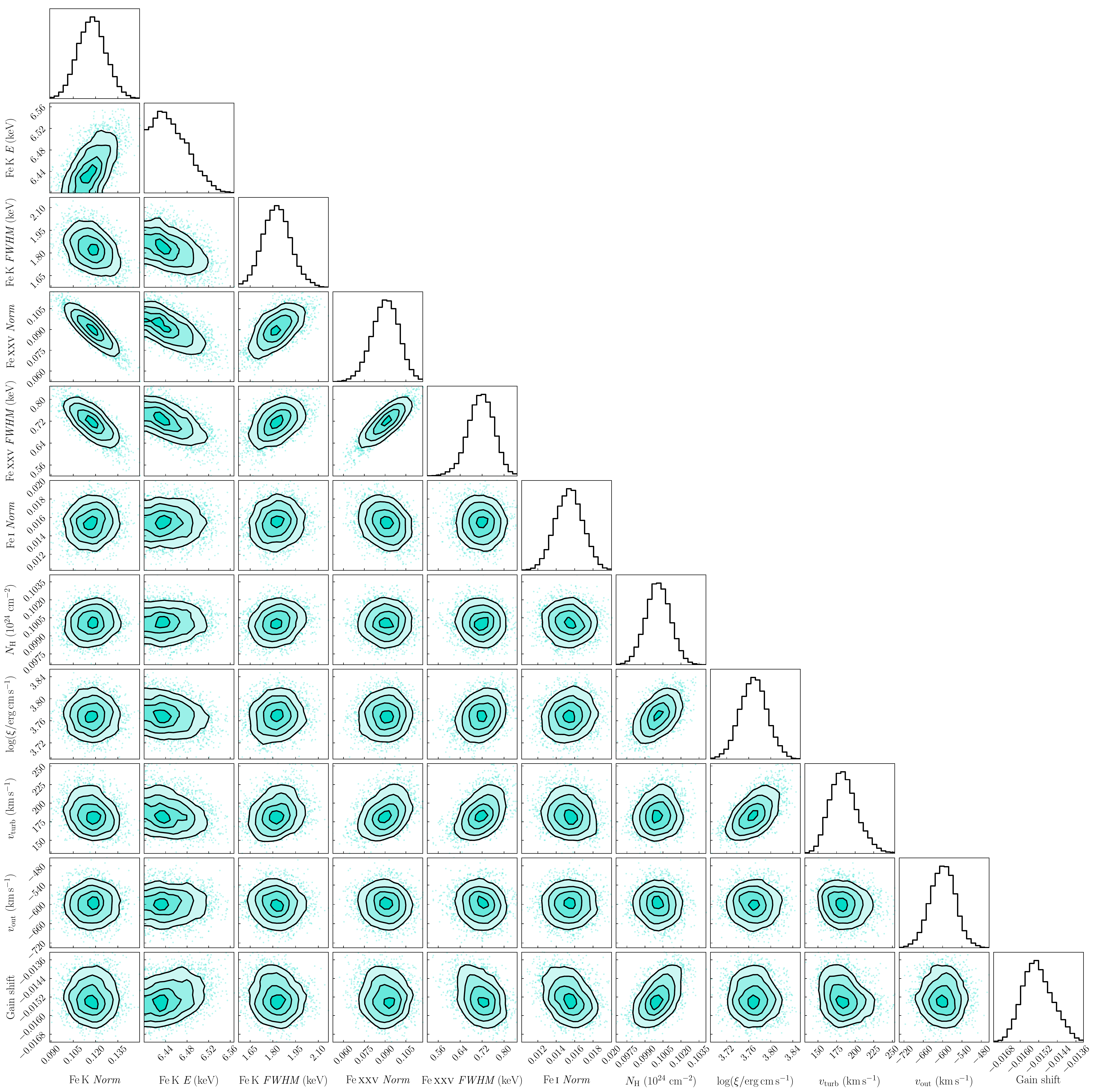}
\caption{\textbf{Parameter posterior plot of observation 0865440101.} Corner plot showing 1- and 2-dimensional marginalized posterior distribution for a subset of the free parameters of the continuum + disk wind model, from observation 0865440101 (segments 2-6 stacked). The analysis includes 6 Fe K band emission line parameters, 4 disk wind parameters, and 1 gain shift parameter. The contours show 0.5, 1, 1.5 and 2$\sigma$ confidence levels.
\label{MCMCfig}}
\end{center}
\end{figure}

\begin{figure*}
\begin{center}
\includegraphics[width=0.7\textwidth]{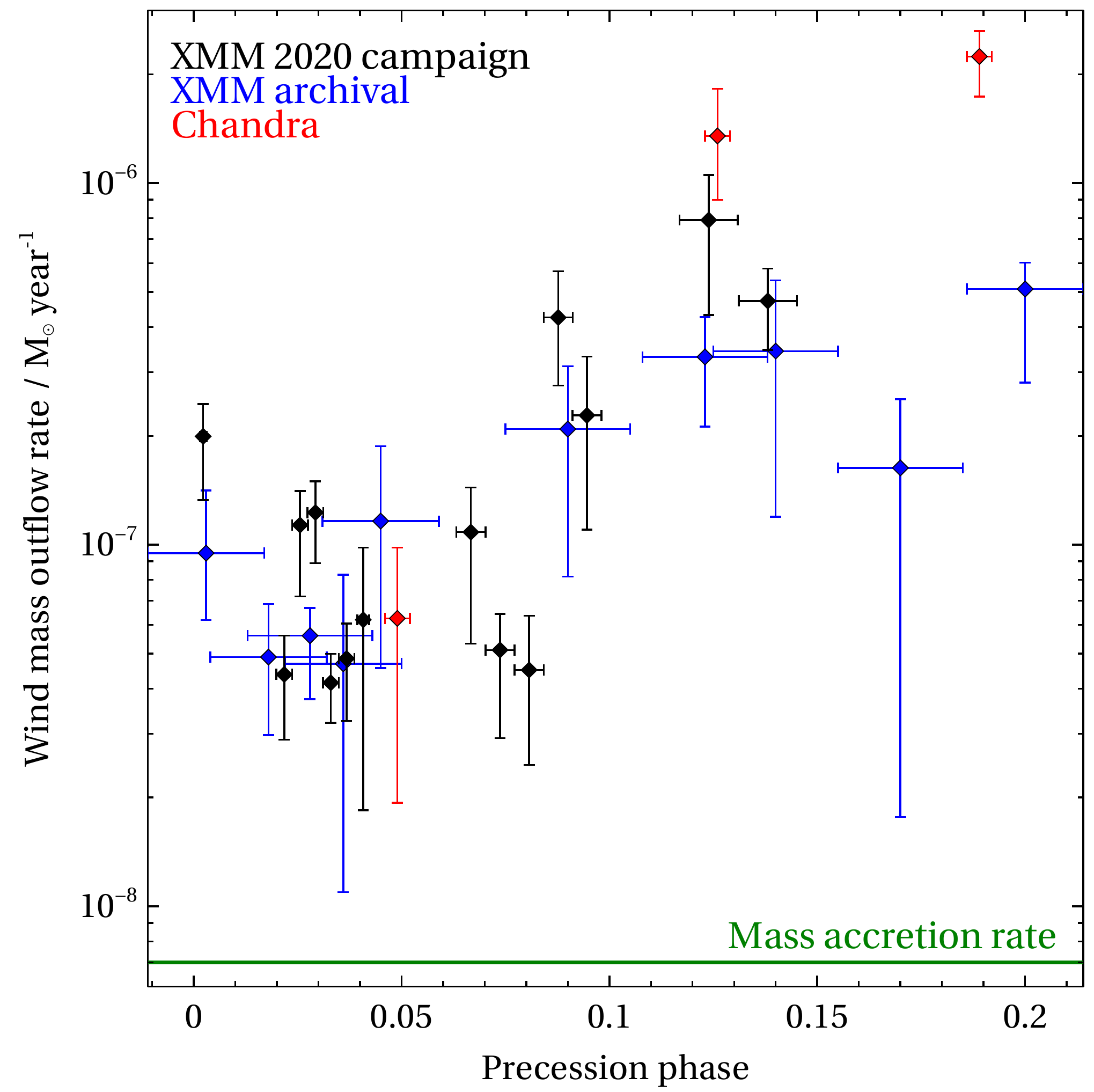}
\caption{\textbf{Wind mass outflow rate.} The wind mass outflow rate calculated by assuming an outflow with a large relative thickness ($\Delta R/R=1$), with a launch solid angle of $4\pi$. Observations from the August 2020 and archival \xmm\ observations are in black and blue colors, \chandra\ observations are in red. The green horizontal line shows the measured mass accretion rate through the outer accretion disk \cite{Boroson+07}. \label{absdist_mass_out}}
\end{center}
\end{figure*}

\begin{table}
\small
\renewcommand{\arraystretch}{1.17}
\centering
\begin{tabular}{cccccccc}
\hline
Observation & Seg.   & Precession & \delcstat & Column  & Ionization  & Outflow  & Velocity \\
 &  & phase &  & density &parameter &  velocity & width \\
 &  & & & 10$^{22}$ \pcm & \logxi & km/s & km/s \\
\hline
\hline

\multicolumn{8}{c}{August 2020 \xmm\ data}\\
\hline
0865440101 & 1 & $0.0023 \pm 0.0008$ & 55.8 & $ 3.7^{+0.7}_{-0.5} $ & $ 2.93^{+0.09}_{-0.06} $ & $ -510^{+90}_{-100} $ & $ 110^{+30}_{-20} $ \\
 & 2 & $0.0218 \pm 0.0019$ & 134.8 & $ 9.1^{+3.5}_{-2.4} $ & $ 3.71^{+0.08}_{-0.10} $ & $ -360^{+70}_{-70} $ & $ 210^{+40}_{-30} $ \\
 & 3 & $0.0256 \pm 0.0019$ & 61.7 & $ 6.3^{+1.8}_{-1.5} $ & $ 3.60^{+0.10}_{-0.09} $ & $ -760^{+150}_{-110} $ & $ 80^{+30}_{-20} $ \\
 & 4 & $0.0293 \pm 0.0019$ & 93.4 & $ 3.6^{+2.0}_{-0.6} $ & $ 3.50^{+0.07}_{-0.07} $ & $ -620^{+90}_{-110} $ & $ 300^{+90}_{-70} $ \\
 & 5 & $0.0330 \pm 0.0019$ & 119.6 & $ 13^{+7}_{-3} $ & $ 3.92^{+0.03}_{-0.03} $ & $ -460^{+70}_{-90} $ & $ 200^{+30}_{-30} $ \\
 & 6 & $0.0368 \pm 0.0019$ & 146.7 & $ 16^{+2}_{-4} $ & $ 3.87^{+0.06}_{-0.09} $ & $ -500^{+90}_{-80} $ & $ 190^{+40}_{-30} $ \\
 & 7 & $0.0408 \pm 0.0014$ & 68.9 & $ 2.1^{+1.1}_{-0.4} $ & $ 3.38^{+0.08}_{-0.08} $ & $ -240^{+150}_{-140} $ & $ 480^{+160}_{-120} $ \\
0865440401 & 1 & $0.0667 \pm 0.0035$ & 91.2 & $ 0.77^{+0.42}_{-0.19} $ & $ 3.20^{+0.07}_{-0.06} $ & $ -220^{+90}_{-70} $ & $ 200^{+100}_{-50} $ \\
 & 2 & $0.0737 \pm 0.0035$ & 30.4 & $ 4.6^{+25.4}_{-0.9} $ & $ 3.93^{+0.09}_{-0.07} $ & $ -570^{+170}_{-120} $ & $ 120^{+180}_{-50} $ \\
 & 3 & $0.0807 \pm 0.0035$ & 29.5 & $ 4.6^{+2.0}_{-2.1} $ & $ 3.95^{+0.11}_{-0.08} $ & $ -530^{+160}_{-200} $ & $ 270^{+200}_{-130} $ \\
 & 4 & $0.0877 \pm 0.0035$ & 20.5 & $ 0.20^{+0.57}_{-0.07} $ & $ 3.02^{+0.08}_{-0.06} $ & $ -580^{+130}_{-190} $ & $ 210^{+140}_{-130} $ \\
 & 5 & $0.0946 \pm 0.0035$ & 21.1 & $ 0.37^{+0.42}_{-0.13} $ & $ 3.24^{+0.14}_{-0.09} $ & $ -510^{+190}_{-220} $ & $ 290^{+220}_{-180} $ \\
0865440501 & 1 & $0.1239 \pm 0.0070$ & 7.6 & $ 0.050^{+0.058}_{-0.023} $ & $ 2.89^{+0.11}_{-0.05} $ & $ -820^{+240}_{-260} $ & $ 190^{+270}_{-160} $ \\
 & 2 & $0.1381 \pm 0.0070$ & 24.8 & $ 0.078^{+0.038}_{-0.017} $ & $ 2.85^{+0.04}_{-0.04} $ & $ -470^{+90}_{-100} $ & $ 100^{+110}_{-50} $ \\
\hline
\multicolumn{7}{c}{Archival \xmm\ data}\\
\hline
0134120101 & & $0.170 \pm 0.015 $  & 6.9 & $ 0.17^{+0.10}_{-0.08} $ & $ 2.93^{+0.14}_{-0.07} $ & $ -270^{+230}_{-140} $ & $ 170^{+200}_{-110} $ \\
0153950301 & & $0.036 \pm 0.014 $  & 29.3 & $ 1.9^{+1.6}_{-0.9} $ & $ 3.58^{+0.18}_{-0.17} $ & $ -230^{+150}_{-160} $ & $ 230^{+160}_{-100} $ \\
0673510501 & & $0.028 \pm 0.015 $  & 66.0 & $ 26^{+20}_{-13} $ & $ 4.01^{+0.09}_{-0.07} $ & $ -780^{+130}_{-100} $ & $ 100^{+30}_{-30} $ \\
0673510601 & & $0.123 \pm 0.015 $  & 28.3 & $ 0.25^{+0.27}_{-0.09} $ & $ 3.13^{+0.06}_{-0.07} $ & $ -500^{+130}_{-130} $ & $ 150^{+180}_{-80} $ \\
0673510801 & & $0.104 \pm 0.015 $  & 11.5 & $ 0.34^{+0.55}_{-0.21} $ & $ 3.05^{+0.22}_{-0.17} $ & $ 170^{+650}_{-710} $ & $ 1800^{+1060}_{-740} $ \\
0673510901 & & $0.090 \pm 0.015 $  & 16.5 & $ 0.24^{+0.16}_{-0.10} $ & $ 2.96^{+0.13}_{-0.09} $ & $ -300^{+140}_{-140} $ & $ 130^{+120}_{-80} $ \\
0783770501 & & $0.003 \pm 0.014 $  & 45.0 & $ 4.6^{+0.8}_{-1.4} $ & $ 2.92^{+0.05}_{-0.04} $ & $ -250^{+70}_{-120} $ & $ 70^{+20}_{-20} $ \\
0783770601 & & $0.018 \pm 0.014 $  & 94.4 & $ 14^{+32}_{-6} $ & $ 3.68^{+0.10}_{-0.16} $ & $ -410^{+80}_{-90} $ & $ 180^{+60}_{-30} $ \\
0783770701 & & $0.045 \pm 0.014 $  & 28.2 & $ 2.5^{+3.2}_{-1.3} $ & $ 3.59^{+0.20}_{-0.26} $ & $ -660^{+230}_{-280} $ & $ 610^{+470}_{-370} $ \\
0830530101 & & $0.200 \pm 0.014 $  & 9.0 & $ 0.056^{+0.039}_{-0.026} $ & $ 2.67^{+0.04}_{-0.04} $ & $ -410^{+160}_{-70} $ & $ 20^{+20}_{-10} $ \\
0830530401 & & $0.140 \pm 0.015 $  & 26.2 & $ 1.5^{+2.2}_{-0.6} $ & $ 3.27^{+0.25}_{-0.20} $ & $ -1180^{+400}_{-510} $ & $ 1490^{+780}_{-480} $ \\
\hline
\multicolumn{8}{c}{\chandra\ data}\\
\hline
2704 & & $0.189 \pm 0.003$ & 27.3 & $ 0.015^{+0.003}_{-0.004} $ & $ 2.42^{+0.06}_{-0.07} $ & $ -800^{+90}_{-90} $ & $ 250^{+70}_{-70} $ \\
23356 & & $0.049 \pm 0.003$ & 38.7 & $ 1.4^{+0.8}_{-0.5} $ & $ 3.67^{+0.23}_{-0.16} $ & $ -310^{+140}_{-150} $ & $ 480^{+190}_{-210} $ \\
23360 & & $0.126 \pm 0.003$ & 13.3 & $ 0.021^{+0.007}_{-0.007} $ & $ 2.54^{+0.11}_{-0.12} $ & $ -590^{+80}_{-160} $ & $ 100^{+230}_{-90} $ \\
\hline
\end{tabular}
\caption{\textbf{Best-fitting disk wind parameters for all observations or segments.} The observation name and segment number (for any observations split into smaller segments) are in the first two columns, the third column contains the calculated precession phase. The fourth column lists the fit improvement \delcstat\ obtained by adding the disk wind model to the continuum-only fit. The following columns list the column density, ionization parameter, orbital motion-corrected outflow velocity and the velocity width.}
\label{resulttable} 
\end{table}

\begin{table}
\small
\renewcommand{\arraystretch}{1.17}
\centering
\begin{tabular}{ccccc}
\hline
Observation & Seg.   & Distance from & Distance from & Height above  \\
 &  & X-ray source & X-ray source & disk \\
 &  & $\Delta R/R \sim 1$ & $n\sim~$const & $n\sim~$const \\
 &  & cm & cm & cm \\
\hline
\hline

\multicolumn{5}{c}{August 2020 \xmm\ data}\\
\hline
0865440101 & 1 & $ 2.65E+11^{+4.90E+10}_{-8.43E+10} $ & $ 1.66E+10^{+1.12E+09}_{-1.65E+09} $ & $ 4.48E+09^{+3.34E+08}_{-4.87E+08} $ \\
 & 2 &		 $ 3.42E+10^{+1.13E+10}_{-1.41E+10} $ & $ 9.29E+09^{+9.03E+08}_{-9.25E+08} $ & $ 2.29E+09^{+2.40E+08}_{-2.43E+08} $ \\
 & 3 &		 $ 5.92E+10^{+1.75E+10}_{-2.25E+10} $ & $ 1.02E+10^{+9.74E+08}_{-1.14E+09} $ & $ 2.51E+09^{+2.60E+08}_{-2.99E+08} $ \\
 & 4 &		 $ 1.39E+11^{+3.13E+10}_{-6.37E+10} $ & $ 1.17E+10^{+8.52E+08}_{-9.78E+08} $ & $ 2.88E+09^{+2.30E+08}_{-2.61E+08} $ \\
 & 5 &		 $ 1.71E+10^{+3.41E+09}_{-6.73E+09} $ & $ 7.92E+09^{+2.40E+08}_{-3.69E+08} $ & $ 1.86E+09^{+6.03E+07}_{-9.24E+07} $ \\
 & 6 &		 $ 1.55E+10^{+5.04E+09}_{-4.13E+09} $ & $ 8.23E+09^{+7.95E+08}_{-7.96E+08} $ & $ 1.92E+09^{+2.00E+08}_{-1.98E+08} $ \\
 & 7 &		 $ 3.11E+11^{+7.70E+10}_{-1.44E+11} $ & $ 1.35E+10^{+1.08E+09}_{-1.26E+09} $ & $ 3.25E+09^{+2.92e+08}_{-3.33E+08} $ \\
0865440401 & 1 & $ 1.61E+12^{+4.44E+11}_{-7.34E+11} $ & $ 1.85E+10^{+1.20E+09}_{-1.54E+09} $ & $ 4.25E+09^{+3.27E+08}_{-4.10E+08} $ \\
 & 2 &		 $ 4.91E+10^{+1.26E+10}_{-4.30E+10} $ & $ 7.90E+09^{+6.11E+08}_{-8.05E+08} $ & $ 1.57E+09^{+1.34E+08}_{-1.73E+08} $ \\
 & 3 &		 $ 4.65E+10^{+2.28E+10}_{-2.17E+10} $ & $ 7.69E+09^{+6.54E+08}_{-9.59E+08} $ & $ 1.47E+09^{+1.38E+08}_{-1.99E+08} $ \\
 & 4 &		 $ 9.14E+12^{+3.44E+12}_{-7.16E+12} $ & $ 2.26E+10^{+1.50E+09}_{-1.97E+09} $ & $ 4.95E+09^{+4.10E+08}_{-5.24E+08} $ \\
 & 5 &		 $ 3.04E+12^{+1.18E+12}_{-1.95E+12} $ & $ 1.76E+10^{+1.54E+09}_{-2.13E+09} $ & $ 3.53E+09^{+3.78E+08}_{-5.06E+08} $ \\
0865440501 & 1 & $ 4.88E+13^{+2.37E+13}_{-3.14E+13} $ & $ 2.60E+10^{+1.42E+09}_{-3.13E+09} $ & $ 4.90E+09^{+3.64E+08}_{-7.69E+08} $ \\
 & 2 &		 $ 3.27E+13^{+7.67E+12}_{-1.28E+13} $ & $ 2.67E+10^{+9.84E+08}_{-1.29E+09} $ & $ 4.64E+09^{+2.41E+08}_{-3.07E+08} $ \\
\hline
\multicolumn{5}{c}{Archival \xmm\ data}\\
\hline
0134120101 & &	 $ 8.80E+12^{+4.33E+12}_{-4.80E+12} $ & $ 2.06E+10^{+1.48E+09}_{-3.21E+09} $ & $ 2.45E+09^{+2.61E+08}_{-5.23E+08} $ \\
0153950301 & &	 $ 2.71E+11^{+1.50E+11}_{-1.75E+11} $ & $ 1.21E+10^{+1.99E+09}_{-2.38E+09} $ & $ 2.91E+09^{+5.36E+08}_{-6.21E+08} $ \\
0673510501 & &	 $ 7.10E+09^{+3.87E+09}_{-3.92E+09} $ & $ 7.10E+09^{+5.04E+08}_{-7.46E+08} $ & $ 1.69E+09^{+1.27E+08}_{-1.87E+08} $ \\
0673510601 & &	 $ 6.62E+12^{+2.61E+12}_{-3.84E+12} $ & $ 2.15E+10^{+1.51E+09}_{-1.55E+09} $ & $ 3.84E+09^{+3.57E+08}_{-3.55E+08} $ \\
0673510801 & &	 $ 6.90E+12^{+4.82E+12}_{-5.51E+12} $ & $ 2.57E+10^{+4.28E+09}_{-7.06E+09} $ & $ 5.38E+09^{+1.20E+09}_{-1.80E+09} $ \\
0673510901 & &	 $ 7.39E+12^{+3.28E+12}_{-4.09E+12} $ & $ 2.22E+10^{+1.99E+09}_{-3.14E+09} $ & $ 4.79E+09^{+5.41E+08}_{-8.16E+08} $ \\
0783770501 & &	 $ 2.07E+11^{+6.52E+10}_{-5.15E+10} $ & $ 1.63E+10^{+6.74E+08}_{-1.00E+09} $ & $ 4.42E+09^{+2.00E+08}_{-2.96E+08} $ \\
0783770601 & &	 $ 2.20E+10^{+1.25E+10}_{-1.70E+10} $ & $ 9.09E+09^{+1.54E+09}_{-1.21E+09} $ & $ 2.26E+09^{+4.14E+08}_{-3.20E+08} $ \\
0783770701 & &	 $ 1.80E+11^{+1.26E+11}_{-1.33E+11} $ & $ 1.11E+10^{+2.46E+09}_{-2.47E+09} $ & $ 2.58E+09^{+6.40E+08}_{-6.18E+08} $ \\
0830530101 & &	 $ 5.56E+13^{+2.67E+13}_{-2.60E+13} $ & $ 2.95E+10^{+1.32E+09}_{-1.43E+09} $ & $ 3.03E+09^{+2.42E+08}_{-2.51E+08} $ \\
0830530401 & &	 $ 4.86E+11^{+2.59E+11}_{-3.83E+11} $ & $ 1.43E+10^{+2.63E+09}_{-3.94E+09} $ & $ 2.01E+09^{+4.81E+08}_{-6.55E+08} $ \\
\hline
\multicolumn{5}{c}{\chandra\ data}\\
\hline
2704 & & 	 $ 4.60E+14^{+1.28E+14}_{-1.32E+14} $ & $ 4.44E+10^{+3.03E+09}_{-3.03E+09} $ & $ 6.99E+09^{+8.69E+08}_{-8.18E+08} $ \\
23356 & &	 $ 3.75E+11^{+1.71E+11}_{-2.41E+11} $ & $ 1.19E+10^{+1.84E+09}_{-2.89E+09} $ & $ 2.75E+09^{+4.77E+08}_{-7.21E+08} $ \\
23360 & &	 $ 2.70E+14^{+1.08E+14}_{-1.13E+14} $ & $ 4.01E+10^{+4.66E+09}_{-4.90E+09} $ & $ 8.89E+09^{+1.54E+09}_{-1.51E+09} $ \\
\hline
\end{tabular}
\caption{\textbf{Calculated wind positions for all observations.} The observation name and segment number (for any observations split into smaller segments) are in the first two columns. The third column contains the calculated maximum distance of the absorber from the X-ray source (assuming $\Delta R/R=1$). The fourth column shows the absorber distance from the X-ray source calculated by assuming constant wind density. The final column contains the height of the absorber above the accretion disk calculated by modelling the warped disk precession and by assuming constant wind density.}
\label{resulttable_calculations} 
\end{table}

\end{document}